\documentclass[preprint,12pt]{elsarticle}

\usepackage{fancyhdr}
\usepackage{fullpage}
\usepackage{amsmath}
\usepackage{amsbsy}
\usepackage{amssymb}
\usepackage{amscd}
\usepackage{amsfonts}
\usepackage{supertabular}
\usepackage{graphics}
\usepackage{verbatim}
\usepackage{epsfig}
\usepackage{xspace}
\usepackage{euscript}
\usepackage{alltt}
\usepackage{boxedminipage}
\usepackage{float}
\usepackage[colorlinks]{hyperref}
\usepackage{color}
\usepackage[all]{xy}
\usepackage{t1enc}
\usepackage{times,exscale}
\usepackage{graphicx,calc}
\usepackage{subfig}
\usepackage[ruled,vlined]{algorithm2e}
\usepackage{epstopdf}
\usepackage{array,multirow}
\usepackage{braket}

\def\R{{\mathbb{R}}}
\def\bx{{\mathbf{x}}}
\def\bR{{\mathbf{R}}}

\def\bH{{\mathbf{H}}}

\def\bw{{\mathbf{w}}}

\journal{arXiv}

\begin{document}

\begin{frontmatter}

%\title{SQDFT: Spectral Quadrature method for $\mathcal{O}(N)$ Kohn-Sham calculations at high temperature}
\title{SQDFT: Spectral Quadrature method for large-scale parallel $\mathcal{O}(N)$ Kohn-Sham calculations at high temperature}
\author[gatech]{Phanish Suryanarayana\corref{cor}}
\author[gatech]{Phanisri P. Pratapa}
\author[gatech]{Abhiraj Sharma}
\author[llnl]{John E. Pask}
\address[gatech]{College of Engineering, Georgia Institute of Technology, Atlanta, GA 30332, USA}
\address[llnl]{Physics Division, Lawrence Livermore National Laboratory, Livermore, CA 94550, USA}
\cortext[cor]{Corresponding Author (\it phanish.suryanarayana@ce.gatech.edu) }

\begin{abstract}
We present SQDFT: a large-scale parallel implementation of the Spectral Quadrature (SQ) method for $\mathcal{O}(N)$ Kohn-Sham Density Functional Theory (DFT) calculations at high temperature.
%We present an $\mathcal{O}(N)$ implementation of the Spectral Quadrature (SQ) method for performing large-scale Kohn-Sham Density Functional Theory calculations at high temperature, a framework referred to as SQDFT. 
Specifically, we develop an efficient and scalable finite-difference implementation of the infinite-cell Clenshaw-Curtis SQ approach, in which results for the infinite crystal are obtained by expressing quantities of interest as bilinear forms or sums of bilinear forms, that are then approximated by spatially localized Clenshaw-Curtis quadrature rules. We demonstrate the accuracy of SQDFT by showing systematic convergence of energies and atomic forces with respect to SQ parameters to reference diagonalization results, and convergence with discretization to established planewave results, for both metallic and insulating systems. We further demonstrate that SQDFT achieves excellent strong and weak parallel scaling on computer systems consisting of tens of thousands of processors, with near perfect $\mathcal{O}(N)$ scaling with system size and wall times as low as a few seconds per self-consistent field iteration. Finally, we verify the accuracy of SQDFT in large-scale quantum molecular dynamics simulations of aluminum at high temperature.
\end{abstract}


\end{frontmatter}
%%%%%%%%%%%%%%%%%%%%%%%%%%%%%%%%%%%%%%%%%%%%%%%%%%%%%%%%%%%%%%%%%%%%%%%%%%%%%%%%%%%%%%%%%%%%%%%%%%%%%%%%%%%%%%%%%%
%%%%%%%%%%%%%%%%%%%%%%%%%%%%%%%%%%%%%%%%%%%%%%%%%%%%%%%%%%%%%%%%%%%%%%%%%%%%%%%%%%%%%%%%%%%%%%%%%%%%%%%%%%%%%%%%%%
\section{Introduction} \label{Sec:Introduction} 
Kohn-Sham Density Functional Theory (DFT) \cite{Hohenberg,Kohn1965} is a powerful tool for predicting and understanding a wide range of materials properties, from the first principles of quantum mechanics, with no empirical or adjustable parameters. The tremendous popularity of DFT is a consequence of its high accuracy to cost ratio relative to other such ab initio theories. However, the solution of the Schr\"odinger type eigenproblem for the Kohn-Sham orbitals remains a challenging task. In particular, since the orbitals need to be orthogonal and increase in number linearly with the number of atoms $N$, the overall computational complexity of DFT calculations scales as $\mathcal{O}(N^3)$ and the memory requirement scales as $\mathcal{O}(N^2)$ (see, e.g., \cite{Goedecker,Bowler2012,aarons2016perspective}).
%\footnote{The scaling associated with both the computational cost and the memory requirement have  large prefactors due to the substantial number of basis functions required for discretization, which increases linearly with system size.} 
The orthogonality constraint on the orbitals also results in global communications between processors in parallel computing, which limits parallel scalability. The need for high performance parallel computing is especially crucial for quantum molecular dynamics (QMD) calculations \cite{marx2009ab,kresse1993ab}, wherein tens or hundreds of thousands of Kohn-Sham solutions can be required to complete a single simulation. 

In order to overcome the critical $\mathcal{O}(N^3)$ scaling bottleneck, much research in the past two decades has been devoted to the development of $\mathcal{O}(N)$ solution strategies (see, e.g., \cite{Goedecker,Bowler2012,aarons2016perspective} and references therein). Rather than calculate the orthonormal Kohn-Sham orbitals, these techniques directly determine the electron density, energy, and atomic forces in $\mathcal{O}(N)$ operations by exploiting the decay of the density matrix \cite{goedecker1998decay,ismail1999locality,zhang2001properties,taraskin2002spatial,benzi2013decay}.\footnote{The real-space density matrix has exponential decay for insulating systems as well as metallic systems at finite temperature \cite{goedecker1998decay,benzi2013decay}.} These efforts have yielded significant advances, culminating in mature implementations of a number of approaches \cite{SIESTA,SIESTAweb,Conquestref,Conquestweb,ONETEP,ONETEPweb,FEMTECKref,MGMolref,BigDFTref,BigDFTweb,OpenMXweb,FreeONref,FreeONweb}. However, significant challenges remain. In particular, the accuracy and stability of $\mathcal{O}(N)$ methods remain ongoing concerns due to the need for additional computational parameters, subtleties in determining sufficient numbers and/or centers of localized orbitals, limitations of underlying basis sets, and calculation of accurate atomic forces, as required for structural relaxation and molecular dynamics simulations \cite{Bowler2012,RuiHinSky12}. In addition, efficient large-scale parallelization poses a significant challenge due to complex communications patterns and load balancing issues. Finally, and perhaps most importantly, the assumption of a band gap in the electronic structure makes existing methods inapplicable to metallic systems \cite{Bowler2012,aarons2016perspective}.\footnote{This is also the case for insulating systems at sufficiently high temperature which results in conduction bands becoming partially occupied.} 

%In this work, we develop a framework for large-scale Kohn-Sham calculations at high temperature, applicable to metals and insulators alike. Such calculations have a number of applications, including the study of warm dense matter---a state encountered in high-power laser experiments, the interiors of planets, and the crusts of white dwarfs and neutron stars \cite{renaudin2003aluminum,dharma2006static,ernstorfer2009formation,white2013orbital,pribram2014thermal,cangi2015efficient}. 
%However, high-temperature DFT calculations 
%%---the electronic temperature/smearing given by the Fermi-Dirac distribution function is set to be equal to the ionic temperature---
%present unique challenges for $\mathcal{O}(N^3)$ methods as well as local-orbital based $\mathcal{O}(N)$ methods. These include the need for a significantly larger number of orbitals to be calculated, as the number of partially occupied states increases, and need for more diffuse orbitals, as higher-energy states become less localized. Consequently, $\mathcal{O}(N^3)$ methods as well as local-orbital based $\mathcal{O}(N)$ methods are associated with extremely large prefactors, which makes high-temperature QMD calculations for even small systems currently intractable. 

High-temperature DFT calculations present additional challenges \cite{gradesred2014,grabasben2011}. 
Such calculations have a number of applications, including the study of warm dense matter and dense plasmas, as occur in laser experiments, and the interiors of giant planets and stars \cite{gradesred2014,grabasben2011,renaudin2003aluminum,dharma2006static,ernstorfer2009formation,white2013orbital}. 
%However, high-temperature DFT calculations  
%---the electronic temperature/smearing given by the Fermi-Dirac distribution function is set to be equal to the ionic temperature---
%present unique challenges for $\mathcal{O}(N^3)$ methods as well as local-orbital based $\mathcal{O}(N)$ methods. 
Particular challenges include the need for a significantly larger number of orbitals to be computed, as the number of partially occupied states increases, and need for more diffuse orbitals, as higher-energy states become less localized. Consequently, $\mathcal{O}(N^3)$ methods as well as local-orbital based $\mathcal{O}(N)$ methods have very large prefactors, which makes high-temperature QMD calculations for even small systems intractable. Recent work to address these challenges includes orbital-free molecular dynamics (OFMD) \cite{lamclegil2006} wherein the standard Kohn-Sham kinetic energy is replaced by an approximation in terms of the density, extended first principles molecular dynamics (ext-FPMD) \cite{zhang2016extended} wherein higher-energy states are approximated as planewaves rather than computed explicitly, and finite-temperature potential functional theory (PFT) \cite{cangi2015efficient} wherein an orbital-free free energy approximation is constructed through a coupling-constant formalism. While OFMD can miss electronic shell structure effects \cite{humilkre2016}, ext-FPMD and PFT have been shown to capture such effects in initial applications.

The recently developed Spectral Quadrature (SQ) method for $\mathcal{O}(N)$ Kohn-Sham calculations \cite{suryanarayana2013spectral,pratapa2016spectral} addresses both scaling with number of atoms and scaling with temperature, while retaining systematic convergence to standard $\mathcal{O}(N^3)$ results for metals and insulators alike. 
In this approach, all quantities of interest are expressed as bilinear forms or sums of bilinear forms, which are then approximated by Clenshaw-Curtis quadrature rules that remain spatially localized by exploiting the locality of electronic interactions in real-space \cite{prodan2005nearsightedness}, i.e, the exponential decay of the density matrix in real-space for insulators as well as metals at finite temperature. In conjunction with local reformulation of the electrostatics, this technique enables the $\mathcal{O}(N)$ evaluation of the electronic density, energy, and atomic forces. The computational cost of SQ decreases rapidly with increasing temperature due to the enhanced locality of the electronic interactions and the increased smoothness of the Fermi-Dirac function. Further, it is well suited to scalable high-performance parallel computing since a majority of the communication is localized to nearby processors, whose pattern remains fixed throughout the simulation. The SQ approach also permits infinite-crystal calculations without recourse to Brillouin zone integration or large supercells, a technique referred to as the infinite-cell method \cite{pratapa2016spectral}.

In this paper, we present SQDFT: a parallel implementation of the SQ method for $\mathcal{O}(N)$ Kohn-Sham DFT calculations at high temperature.\footnote{Though we focus on high-temperature calculations in this work, SQDFT is also capable of performing $\mathcal{O}(N)$ DFT calculations at ambient temperature, with a larger prefactor (Appendix \ref{App:LowT}).} Specifically, we develop a finite-difference implementation of the  infinite-cell variant of the Clenshaw-Curtis SQ method that can efficiently scale on large-scale parallel computers. 
%In order to verify the accuracy of SQDFT, we ensure systematic convergence of energy and atomic forces with respect to the SQ parameters to reference diagonalization results; and their convergence with discretization to well-established planewave results. We also demonstrate that SQDFT achieves excellent strong and weak scaling on computer systems consisting of tens of thousands of cores, having close to perfect $\mathcal{O}(N)$ scaling with system size and wall times as low as a few seconds per self-consistent field (SCF) iteration. Finally, we verify the capability of SQDFT to perform high-temperature QMD simulations.
We verify the accuracy of SQDFT by showing systematic convergence of energies and atomic forces to reference diagonalization results, and convergence with discretization to established planewave results, for both metallic and insulating systems. We further show that SQDFT achieves excellent strong and weak parallel scaling on computer systems consisting of tens of thousands of cores, with near perfect $\mathcal{O}(N)$ scaling with system size and wall times as low as a few seconds per self-consistent field (SCF) iteration. Finally, we verify the accuracy of SQDFT in large-scale quantum molecular dynamics simulations of aluminum at high temperature.

The remainder of this paper is organized as follows. In Section \ref{Sec:DFT}, we review the $\mathcal{O}(N)$ density matrix formulation of DFT. Next, we discuss the formulation and implementation of SQDFT in Section \ref{Sec:SQDFT} and study its accuracy, efficiency, and scaling in Section \ref{Sec:Results}. Finally, we provide concluding remarks in Section \ref{Sec:Conclusions}.

%%%%%%%%%%%%%%%%%%%%%%%%%%%%%%%%%%%%%%%%%%%%%%%%%%%%%%%%%%%%%%%%%%%%%%%%%%%%%%%%%%%%%%%%%%%%%%%%%%%%%%%%%%%%%%%%%%%
\section{$\mathcal{O}(N)$ Density Functional Theory} \label{Sec:DFT}
Consider a cuboidal domain $\Omega$ containing $N$ atoms, the unit cell of an infinite crystal. Let the nuclei be positioned at $\bR = \{\bR_1, \bR_2, \ldots, \bR_N \}$ and let there be a total of $N_e$ valence electrons. Neglecting spin and Brillouin zone integration, the nonlinear eigenproblem for the electronic ground state in Kohn-Sham Density Functional Theory (DFT) can be written as \cite{pratapa2016spectral,anantharaman2009existence}
\begin{eqnarray}
\mathcal{D} & = & g(\mathcal{H},\mu,\sigma) = \left( 1 + \exp \left(\frac{\mathcal{H}-\mu \mathcal{I}}{\sigma} \right)   \right)^{-1} \,, \label{Eqn:SelfCons:DensityMatrix} \\
\mathcal{H} & = & -\frac{1}{2} \nabla^2 + V_{xc} + \phi + \mathcal{V}_{nl} \,, \label{Eqn:Hamiltonian}
\end{eqnarray} 
where $\mathcal{D}$ is the density matrix; $g$ is the Fermi-Dirac function; $\mu$ is the Fermi level, which is determined by solving for the constraint on the total number of electrons, 
i.e., $2 \mathrm{Tr}(\mathcal{D}) = N_{e}$; 
%($\mathrm{Tr}$ denotes the trace)
$\sigma = k_{B} T$ is the smearing, where $k_{B}$ is Boltzmann's constant and $T$ is the electronic temperature\footnote{The electronic temperature/smearing is typically set to be equal to the ionic temperature in QMD simulations, particularly for those performed at high temperature.}; $\mathcal{H}$ is the Hamiltonian; $V_{xc}$ is the exchange-correlation potential; $\mathcal{V}_{nl}$ is the nonlocal pseudopotential; and $\phi$ is the electrostatic potential, the solution to the Poisson equation \cite{Pask2005,Gavini2007,Phanish2010,Phanish2011}
\begin{equation} \label{Eqn:PoissonEqn}
-\frac{1}{4 \pi} \nabla^2 \phi(\bx,\bR) = \rho_{\mathcal{D}}(\bx) + b(\bx,\bR) 
\end{equation}     
subject to periodic boundary conditions. Above, 
\begin{equation}
\rho_{\mathcal{D}}(\bx) = 2 \mathcal{D}(\bx,\bx) \label{Eq:rho}
\end{equation}
is the electron density and $b = \sum_{I} b_I$ is the total pseudocharge density of the nuclei \cite{Pask2005}, where $b_{I}$ is the pseudocharge density of the $I^{th}$ nucleus and the index $I$ extends over all atoms in $\R^3$.

Once the electronic ground state has been determined, the free energy can be written as \cite{pratapa2016spectral}\footnote{The repulsive energy correction for overlapping pseudocharges \cite{Suryanarayana2014524,ghosh2016higher} has been explicitly included in the free energy expression. This term plays a particularly important role in high-temperature simulations since the ions get significantly closer compared to ambient temperature. \label{Footnote:Ecorr}} 
\begin{eqnarray} 
\mathcal{F} (\bR) & = & 2\text{Tr}(\mathcal{D} \mathcal{H})  + E_{xc}(\rho_{\mathcal{D}}) - \int_{\Omega} V_{xc}(\rho_{\mathcal{D}}(\bx)) \rho_{\mathcal{D}}(\bx) \, \mathrm{d\bx} + \frac{1}{2} \int_{\Omega} (b(\bx,\bR)-\rho_{\mathcal{D}}(\bx)) \phi(\bx,\bR) \, \mathrm{d\bx} \nonumber \\
& +  & \frac{1}{2} \int_{\Omega} \left( \tilde{b}(\bx,\bR) + b(\bx,\bR) \right) V_{c} (\bx,\bR) \, \mathrm{d\bx} - \frac{1}{2}\sum_{I} \int_{\Omega} \tilde{b}_I(\bx,\bR_I) \tilde{V}_I(\bx,\bR_I) \, \mathrm{d\bx} \nonumber \\
& + & 2 \sigma \text{Tr}\left( \mathcal{D} \log \mathcal{D} + (\mathcal{I}-\mathcal{D}) \log (\mathcal{I}-\mathcal{D}) \right) \,, \label{DM:groundstate}
\end{eqnarray}
where $E_{xc}$ is the exchange-correlation energy; $\tilde{b} = \sum_{I} \tilde{b}_I $ is the total reference pseudocharge density,  where $\tilde{b}_{I}$ is the reference pseudocharge density of the $I^{th}$ nucleus and the summation index $I$ runs over all atoms in $\R^3$; $V_c = \sum_{I} V_{c,I} = \sum_{I} \left(\tilde{V}_I-V_I \right)$, where $\tilde{V}_I$ and $V_I$ are the potentials generated by $\tilde{b}_I$ and $b_I$, respectively; and $\mathcal{I}$ is the identity operator. The first term is the band structure energy ($E_{band}$) and the last term is the energy associated with the electronic entropy ($E_{ent}$). 

In order to update the positions of the ions during the Born-Oppenheimer quantum molecular dynamics (QMD) simulations, the Hellmann-Feynman force on the $I^{th}$ nucleus can be written as
\cite{pratapa2016spectral}\footnote{For the reasons mentioned in footnote \ref{Footnote:Ecorr}, the force corresponding to the repulsive energy correction for overlapping pseudocharges \cite{Suryanarayana2014524,ghosh2016higher} has been explicitly included in the atomic force expression. Note that this term has been significantly simplified from its original form by utilizing the relation $\int \nabla f(|\bx-\bR_I|) g(|\bx-\bR_I|) \, \mathrm{d\bx}=0$ for spherically symmetric functions $f$ and $g$, which in the present case are $b_I$, $b_{I'}$, $V_I$, and $V_{I'}$.}
\begin{eqnarray}  
\mathbf{f}_{I} & = & \sum_{I'} \int_{\Omega} \nabla b_{I'}(\bx,\bR_{I'}) \phi(\bx,\bR) \, \mathrm{d\bx} +  \frac{1}{2} \sum_{I'} \int_{\Omega} \bigg[ \left(\tilde{b}(\bx,\bR)+b(\bx,\bR) \right) \nabla V_{c,I'}  \nonumber \\ 
& + & \left( \nabla \tilde{b}_{I'}(\bx,\bR_{I'}) + \nabla b_{I'}(\bx,\bR_{I'}) \right) V_c(\bx,\bR)   \bigg] \,\mathrm{d\bx} - 4 \sum_{I'} \text{Tr}\left(\mathcal{V}_{nl}^{I'} \nabla \mathcal{D} \right) \,, \label{Eqn:forces}
\end{eqnarray}
where the summation index $I'$ runs over the $I^{th}$ atom and its periodic images, and $\mathcal{V}_{nl,I'}$ is the nonlocal pseudopotential associated with the ${I'}^{th}$ atom. The first two terms together constitute the local component of the force ($\mathbf{f}_I^{l}$) and the last term is the nonlocal component of the force ($\mathbf{f}_I^{nl}$). 

Within a real-space representation, the density matrix $\mathcal{D}$ has exponential decay for insulators as well as metallic systems at finite electronic temperature/smearing \cite{goedecker1998decay,benzi2013decay}. This decay in the density matrix is exploited by linear-scaling methods through truncation within the calculations to enable $\mathcal{O}(N)$ computation of the electron density, energy, and atomic forces. In doing so, it has been observed that there is exponential convergence in the energy and forces with the size of the truncation region for insulating \cite{bowler2002recent,skylaris2007achieving} as well as metallic systems at finite temperature \cite{suryanarayana2017nearsightedness}. Note that in the above description for DFT, we have employed a local reformulation of the electrostatics to enable $\mathcal{O}(N)$ scaling for the complete Kohn-Sham problem. 
%%%%%%%%%%%%%%%%%%%%%%%%%%%%%%%%%%%%%%%%%%%%%%%%%%%%%%%%%%%%%%%%%%%%%%%%%%%%%%%%%%%%%%%%%%%%%%%%%%%%%%%%%%%%%%%%%%%
\section{Formulation and implementation of SQDFT} \label{Sec:SQDFT}
%In this section, we describe the formulation and parallel implementation of SQDFT---an $\mathcal{O}(N)$ framework for performing large-scale, high-temperature quantum molecular dynamics (QMD) simulations using Kohn-Sham Density Functional Theory (DFT). 
SQDFT is a large-scale parallel implementation of the Clenshaw-Curtis Spectral Quadrature (SQ) method\footnote{The Clenshaw-Curtis variant of SQ is chosen here because it is more efficient compared to its Gauss counterpart \cite{Phanish2012}, particularly in the computation of the nonlocal component of the forces \cite{pratapa2016spectral}.} \cite{suryanarayana2013spectral,pratapa2016spectral} for $\mathcal{O}(N)$ Kohn-Sham density functional calculations. 
In this approach, all quantities of interest (in discrete form) are expressed as bilinear forms or sums of bilinear forms, that are then approximated by spatially localized quadrature rules. The method is identically applicable to insulating and metallic systems and well suited to massively parallel computation. Furthermore, the SQ method becomes more efficient as temperature is increased, since electronic interactions become more localized and the representation of the Fermi-Dirac function becomes more compact \cite{suryanarayana2017nearsightedness}. 

We employ the \emph{infinite-cell} version of the Clenshaw-Curtis SQ method, wherein the results corresponding to the infinite crystal are obtained without recourse to Brillouin zone integration or large supercells \cite{pratapa2016spectral}. Specifically, rather than employ Bloch boundary conditions for the orbitals on $\Omega$, zero-Dirichlet boundary conditions are prescribed at infinity, and the relevant components  of the density matrix for spatial points within $\Omega$ are calculated by utilizing the potential within the truncation region surrounding that point \cite{pratapa2016spectral,suryanarayana2017nearsightedness}. Periodic boundary conditions are retained for the electrostatic potential. Indeed, the infinite-cell SQ approach is equivalent to the standard $\Gamma$-point SQ calculation when the size of the truncation region is smaller than the size of the domain, a situation common in large-scale DFT simulations, particularly those at high temperature. 

We utilize a high-order finite-difference discretization in order to exploit the locality of electronic interactions in real space, enable systematic convergence, and facilitate large-scale parallel implementation. We solve the fixed-point problem in Eq.~\ref{Eqn:SelfCons:DensityMatrix} using the self consistent field (SCF) method \cite{lin2013elliptic}\footnote{In SQDFT, we perform a fixed-point iteration with respect to the effective potential: $V_\text{eff} = V_{xc} + \phi$ .}, whose convergence is accelerated using the Periodic Pulay mixing scheme \cite{banerjee2016periodic}, a technique that significantly outperforms the well-established Anderson/Pulay mixing \cite{anderson1965iterative,pulay1980convergence}. We solve the Poisson problem in Eq.~\ref{Eqn:PoissonEqn} using the Alternating Anderson-Richardson (AAR) method \cite{pratapa2016anderson,suryanarayana2016alternating}, an approach that outperforms the conjugate gradient method \cite{Shewchuk1994} in the context of large-scale parallel computations \cite{suryanarayana2016alternating}. We perform NVE (microcanonical) simulations using the leapfrog method and NVT (canonical) simulations using the Verlet algorithm with the Nose-Hoover thermostat. At each MD step, we extrapolate the electron density using information from the previous two steps \cite{alfe1999ab} to reduce SCF iterations. We parallelize the calculations using domain decomposition, with the communication between processors handled via the Message Passing Interface (MPI) \cite{gropp1999using}.

\begin{figure}[h]
\centering
\includegraphics[keepaspectratio=true,width=0.75\textwidth]{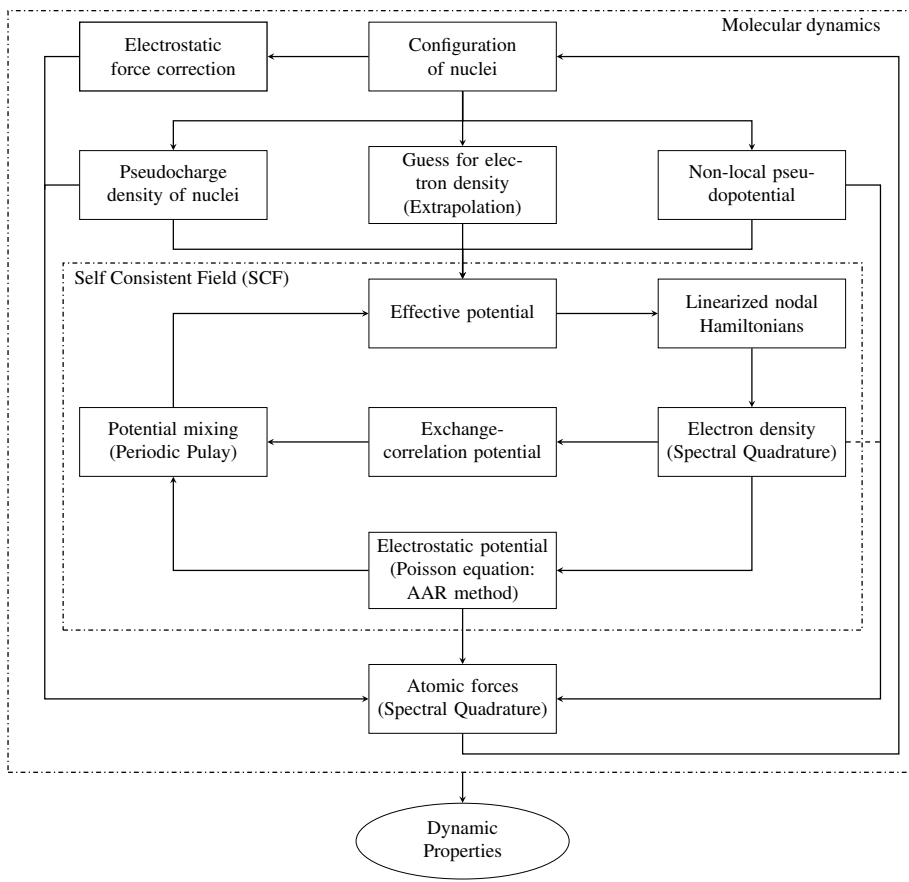}
\caption{Outline of quantum molecular dynamics (QMD) simulation.}
\label{Fig:SQDFT_Flowchart}
\end{figure}

In Fig.~\ref{Fig:SQDFT_Flowchart}, we outline the key steps in a QMD simulation. We employ the Clenshaw-Curtis SQ method for calculating the electron density in each SCF iteration as well as for computing the energy and atomic forces (nonlocal component) once the electronic ground state has been determined. In the sections below, 
%within the framework provided by the finite-difference discretization,
we discuss the calculation of the electron density, energy, and atomic forces in SQDFT. For additional details on the formulation and implementation of the pseudocharges, AAJ/AAR method, and Periodic Pulay mixing scheme, we refer the reader to the relevant previous works  \cite{ghosh2016sparc2,pratapa2016anderson,banerjee2016periodic,suryanarayana2016alternating}.
%%%%%%%%%%%%%%%%%%%%%%%%%%%%%%%%%%%%%%%%%%%%%%%%%%%%%
\subsection{Finite-difference discretization} \label{Subsec:FDDiscretization}
We consider a cubical domain $\Omega$ and discretize it with a uniform grid of spacing $h$, resulting in $N_d = n^3$ finite-difference nodes, the collection of which is referred to as $K_{\Omega}$. We parallelize the calculations by employing domain decomposition, i.e., we partition the domain into cubes of equal size such that $\Omega = \bigcup\limits_{p=1}^{N_P} \Omega_p$, where $N_P$ is the total number of processors and $\Omega_p$ denotes the domain local to the $p^{th}$ processor, each of which contains $N_d^p = N_d/N_P$ finite-difference nodes. We refer to the collection of finite-difference nodes belonging to the $p^{th}$ processor as $K_{\Omega}^{p}$, where $K_{\Omega} = \bigcup\limits_{p=1}^{N_P} K_{\Omega}^p$ with $K_{\Omega}^p \cap K_{\Omega}^q = \emptyset$ if $p \neq q$. 

We approximate the Laplacian arising in the Hamiltonian and the local electrostatic reformulation using the central finite-difference approximation:
\begin{eqnarray}
\nabla^2_h f \big|^{(i,j,k)} & \approx & \sum_{d=0}^{n_o} w_{d} \left(f^{(i+d,j,k)} + f^{(i-d,j,k)} + f^{(i,j+d,k)} + f^{(i,j-d,k)} + f^{(i,j,k+d)} + f^{(i,j,k-d)} \right) \,, \nonumber \\
w_0 & = & - \frac{1}{h^2} \sum_{q=1}^{n_o} \frac{1}{q^2} \,,  \label{Eqn:FD:Laplacian} \\
w_d & = & \frac{2 (-1)^{d+1}}{h^2 d^2} \frac{(n_o!)^2}{(n_o-d)! (n_o+d)!} \,, \,\, d=1, 2, \ldots, n_o \,, \nonumber
\end{eqnarray}
where $f^{(i,j,k)}$ denotes the value of the function $f$ at the node indexed by $(i,j,k)$ and $2n_o$ is the order of the approximation. Similarly, we approximate the gradient operator arising in the atomic forces using central finite-differences:
\begin{eqnarray}
\nabla_h f \big|^{(i,j,k)} & \approx & \sum_{d=1}^{n_o} \tilde{w}_d \bigg( ( f^{(i+d,j,k)} - f^{(i-d,j,k)}) \hat{\mathbf{e}}_1 + ( f^{(i,j+d,k)} - f^{(i,j-d,k)}) \hat{\mathbf{e}}_2  + ( f^{(i,j,k+d)} - f^{(i,j,k-d)}) \hat{\mathbf{e}}_3 \bigg) \,, \nonumber \\
\tilde{w}_d & = & \frac{(-1)^{d+1}}{h d} \frac{(n_o!)^2}{(n_o-d)! (n_o+d)!} \,, \,\, d=1, 2, \ldots, n_o \,, \label{Eqn:FD:Gradient}
\end{eqnarray}
where $\hat{\mathbf{e}}_1$, $\hat{\mathbf{e}}_2$, and $\hat{\mathbf{e}}_3$ are the unit vectors along the edges of $\Omega$. We enforce periodic boundary conditions by mapping any index that does not correspond to a node in the finite-difference grid to its periodic image within $\Omega$. We approximate the spatial integrals arising in the Hamiltonian (projectors of the nonlocal pseudopotential), energy, and atomic forces using the trapezoidal rule:
\begin{equation} \label{Eqn:IntApprox}
\int_{\Omega} f(\bx) \, \mathrm{d\bx} \approx h^3  \sum_{i, j, k=1}^{n} f^{(i,j,k)} \,.
\end{equation}
Even though this is a low order quadrature scheme, it has been chosen to ensure that the discrete free energy is consistent with the discrete Kohn-Sham equations, i.e., the calculated electronic ground state corresponds to the minimum of the free energy within the finite-difference discretization \cite{ghosh2016sparc1,ghosh2016sparc2}. 

In accordance with the nearsightedness principle \cite{prodan2005nearsightedness}, we define the \textit{region of influence} for any finite-difference node as the cube of side $2R_\text{cut}$ centered at that node. A cube is chosen rather than a sphere due to its simplicity and efficiency within the finite-difference implementation. The parameter $R_\text{cut}$ corresponds to the distance beyond which electronic interactions are ignored, i.e., theoretically $R_\text{cut}$ corresponds to the truncation radius for the density matrix. Within the above described finite-difference framework, we define the nodal Hamiltonian $\bH_q \in \R^{N_s \times N_s}$ for any node $q \in K_{\Omega}$ as the restriction of the Hamiltonian to its region of influence \cite{pratapa2016spectral}, where $N_s = (2R_\text{cut}/h + 1)^3$ is the number of finite-difference nodes within the region of influence. Similarly, $\mathbf{w}_q \in \R^{N_s \times 1} $, $\nabla_{h,q} \in \R^{N_s \times N_s}$, and $\mathbf{V}_{nl,q}^I \in \R^{N_s \times N_s}$ represent the restriction of the standard basis vector, gradient matrix, and nonlocal pseudopotential matrix of the $I^{th}$ atom to the region of influence, respectively. It is important to note that $\bH_q$, $\nabla_{h,q}$, and $\mathbf{V}_{nl,q}^{I}$ are not explicitly determined/stored in SQDFT, rather their multiplication with a vector is directly computed in a matrix-free way.

%%%%%%%%%%%%%%%%%%%%%%%%%%%%%%%%%%%%%%%%%%%%%%%%%%%%%
\subsection{Electron density} \label{Subsec:ElectronDensity}
In each iteration of the SCF method, the electron density (Eq.~\ref{Eq:rho}) needs to be computed at all the finite-difference nodes. In SQDFT, the electron density at the $q \in K_{\Omega}^p$ node in the $p^{th}$ processor is calculated using the relations \cite{pratapa2016spectral}
\begin{eqnarray} 
\rho_q & = & \frac{2}{h^3} \sum_{j=0}^{n_{pl}} c_q^j \rho_q^j \,, \label{Eq:ElectronDensity} \\
c_q^j & = & \frac{2}{\pi} \int_{-1}^{1} \frac{g(r,\hat{\mu}_q,\hat{\sigma}_q) T_j(r)}{\sqrt{1-r^2}} \, \mathrm{dr}\,, \quad j=0, 1, \ldots, n_{pl} \,, \label{Eq:cqj}\\
\rho_q^j & = &  
\begin{cases} 
\mathbf{w}_q^T \mathbf{t}_q^j\,, \quad j=0, 1, \ldots, \frac{n_{pl}}{2} \\
\rho_{q}^1 - 2\left(\mathbf{t}_q^{\frac{j+1}{2}} \right)^T \mathbf{t}_{q}^{\frac{j-1}{2}} \,, \quad j=\frac{n_{pl}}{2}+1, \frac{n_{pl}}{2}+3, \ldots, n_{pl}-1 \\
\rho_{q}^0 - 2\left(\mathbf{t}_q^{\frac{j}{2}} \right)^T \mathbf{t}_{q}^{\frac{j}{2}} \,, \quad j=\frac{n_{pl}}{2}+2, \frac{n_{pl}}{2}+4, \ldots, n_{pl}
\end{cases} \label{Eq:rhoqj}
\end{eqnarray}
where $n_{pl}$ is the order of the Clenshaw-Curtis quadrature, chosen here to be a multiple of four for simplicity; $T_j$ denotes the Chebyshev polynomial of degree $j$; $c_q^j$ is the coefficient of $T_j$ in the polynomial expansion of the Fermi-Dirac function, with the value of $c_q^0$ half of that given in the expression; $\hat{\mu}_q=(\mu-\chi_q)/\zeta_q$ is the scaled and shifted Fermi energy, where $\chi_q = (\lambda_q^\text{max}+\lambda_q^\text{min})/2$ and $\zeta_q = (\lambda_q^\text{max}-\lambda_q^\text{min})/2$, with $\lambda_q^\text{max}$ and $\lambda_q^\text{min}$ denoting the maximum and minimum eigenvalues of $\bH_q$, respectively; $\hat{\sigma}_q = \sigma/\zeta_q$ is the scaled smearing; and $\mathbf{t}_q^j = T_j(\hat{\mathbf{H}}_q) \mathbf{w}_q \in \R^{N_s \times 1}$, which is determined using the following iteration, obtained as a consequence of the three term recurrence relation of Chebyshev polynomials:
%\footnote{$T_{j+1}(r) = 2 r T_j(r) - T_{j-1}(r)$.}:
\begin{eqnarray}
\mathbf{t}^{i+1}_q &=&  2\hat{\mathbf{H}}_q \mathbf{t}^{i}_q-\mathbf{t}^{i-1}_q \,, \quad i=1, 2, \ldots, \frac{n_{pl}}{2} \nonumber \\ 
\mathbf{t}_q^1 & = & \hat{\mathbf{H}}_q \mathbf{w}_q \,,\,\,  \mathbf{t}_q^0 = \mathbf{w}_q \,, \label{Eqn:Rec:FD}
\end{eqnarray}
where $\hat{\bH}_q = (\bH_q-\chi_q \mathbf{I})/\zeta_q$ is the scaled and shifted nodal Hamiltonian whose spectrum lies in the interval $[-1,1]$. Note that the iteration in Eq.~\ref{Eqn:Rec:FD} proceeds only up to $n_{pl}/2$ rather than $n_{pl}$, since we have employed the product property of Chebyshev polynomials\footnote{$2 T_{j}(r) T_{k}(r) = T_{j+k}(r) + T_{|j-k|}(r)$. Note that in previous work where the expression for the electron density has been derived \cite{pratapa2016spectral}, this property has not been utilized, and therefore the iteration in Eq.~\ref{Eqn:Rec:FD} proceeds up to $n_{pl}$. The present approach reduces the cost for the computation of $\rho_q^j$ by a factor of two.}
to directly calculate $\rho_q^j$ for $j=n_{pl}/2 +1, n_{pl}/2+2, \ldots, n_{pl}$, as described by Eq.~\ref{Eq:rhoqj}. Since the values of $\rho_q^j$ are independent of the Fermi level $\mu$, they are first computed and stored. Next, $\mu$ is determined by satisfying the constraint on the total number of electrons:
\begin{equation}
2 \sum_{p=1}^{N_P} \sum_{q\in K_{p}} \sum_{j=0}^{n_{pl}} c_q^j \rho_q^j = N_e \,,
\end{equation}
where the values of $c_q^j$ are given by Eq.~\ref{Eq:cqj}. Finally, for the $c_q^j$ corresponding to the Fermi level $\mu$, the electron density is calculated via Eq.~\ref{Eq:ElectronDensity}. 

Since the simulation domain $\Omega$ is cubical and we have employed a uniform finite-difference grid with uniform domain decomposition, the layout of  effective potential values $V_\text{eff}$ required from neighboring processors as part of the nodal Hamiltonians for grid points in each processor is identical. To accomplish this, we utilize the MPI command \texttt{MPI\_Ineighbor\_alltoallv} \cite{gropp2014using} to communicate the required values of $V_\text{eff}$ (i.e., those within the region of influence for grid points in each processor) between processors. Doing so reduces the number of MPI related calls that would otherwise be required in every matrix-vector multiplication. After the communication is complete, for every finite-difference node $q\in K_{\Omega}^{p}$, we first calculate $\lambda_q^\text{max}$ and $\lambda_q^\text{min}$---maximum and minimum eigenvalues of $\bH_q$, respectively---using the Lanczos method \cite{lanczos1950iteration,pratapa2016numerically}. Next, we perform the recursive iteration in Eq.~\ref{Eqn:Rec:FD}, the results of which are used to calculate $\rho_q^j$ as given in Eq.~\ref{Eq:rhoqj}. The matrix-vector multiplications required as part of the Lanczos method and the recursive iteration in Eq.~\ref{Eqn:Rec:FD} are performed in a matrix-free manner. The Chebyshev coefficients $c_q^j$ are calculated using the discrete orthogonality of the Chebyshev polynomials \cite{gil2007numerical}. The Fermi level is determined using Brent's method \cite{press2007numerical}, with Newton-Raphson's method becoming the preferred choice as the temperature is increased. While doing so, the global communication between processors is handled by using the \texttt{MPI\_Allreduce} command. 

During the electron density calculation, the memory costs are dominated by the storage of three vectors during the recursive iteration in Eq.~\ref{Eqn:Rec:FD} ($t_q^{i+1}$, $t_q^{i}$, and $t_q^{i-1}$) and the storage of $\rho_q^j$ for the calculation of the electron density in Eq.~\ref{Eq:ElectronDensity}. Therefore, the memory costs per processor scale as $\mathcal{O}(3 N_s + n_{pl} N_d^p)$. The computational costs are dominated by the the matrix-vector products in the Lanczos iteration and the recursive iteration in Eq.~\ref{Eqn:Rec:FD}. Therefore, the computational cost per processor scales as $\mathcal{O} \left(n_\text{lancz} N_s N_d^p + \frac{1}{2} n_{pl} N_s N_d^p \right)$, where $n_\text{lancz}$ is the number of Lanczos iterations required for determining $\lambda_q^\text{max}$ and $\lambda_q^\text{min}$. Since $N_s$, $n_{pl}$, and $n_\text{lancz}$ remain independent of system size, the overall memory and computational costs scale linearly with the number of finite-difference nodes in the domain $\Omega$, and therefore $\mathcal{O}(N)$ with respect to the number of atoms. 

%%%%%%%%%%%%%%%%%%%%%%%%%%%%%%%%%%%%%%%%%%%%%%%%%%%%%

\subsection{Free energy} \label{Subsec:Energy}
Once the electronic ground state has been determined, i.e., the SCF iteration has converged, the free energy (Eq.~\ref{DM:groundstate}) needs to be calculated. 

\paragraph{Band structure energy} The band structure energy in the Clenshaw-Curtis SQ method \cite{pratapa2016spectral} takes the following form in parallel computations 
\begin{eqnarray}
E_{band} & = & 2\sum_{p=1}^{N_P}\sum_{q\in K_{p}} \sum_{j=0}^{n_{pl}} (\chi_q c_q^j + \zeta_q d_q^j) \rho_q^{j} \,, \label{Eq:Eband}\\
d_q^j & = & \frac{2}{\pi} \int_{-1}^{1} \frac{r g(r,\hat{\mu}_q,\hat{\sigma}_q) T_j(r)}{\sqrt{1-r^2}} \, \mathrm{dr} \,, \label{Eq:d_q}
\end{eqnarray} 
where $d_q^j$ is the coefficient of $T_j$ in the polynomial expansion of the band structure energy function (i.e., $r g(r)$), with the value of $d_q^0$ half of that given in the expression. In addition, $c_q^j$ and $\rho_q^{j}$ are as given in Eqs. \ref{Eq:cqj} and \ref{Eq:rhoqj}, respectively. 

\paragraph{Electronic entropy energy} The electronic entropy energy in the Clenshaw-Curtis SQ approach \cite{pratapa2016spectral} takes the following form in parallel computations 
\begin{eqnarray}
S & = &  2 \sigma \sum_{p=1}^{N_P} \sum_{q\in K_{p}} \sum_{j=0}^{n_{pl}} e_q^j \rho_q^{j} \label{Eq:Entropy}\,, \\
e_q^j & = & \frac{2}{\pi} \int_{-1}^{1} \frac{g(r,\hat{\mu}_q,\hat{\sigma}_q) \log g(r,\hat{\mu}_q,\hat{\sigma}_q) + (1-g(r,\hat{\mu}_q,\hat{\sigma}_q)) \log (1-g(r,\hat{\mu}_q,\hat{\sigma}_q)) T_j(r)}{\sqrt{1-r^2}} \, \mathrm{dr} \,, \label{Eq:e_q}
\end{eqnarray}
where $e_q^j$ is the coefficient of $T_j$ in the polynomial expansion of the electronic entropy energy function (i.e., $g(r) \log g(r) + (1-g(r)) \log (1-g(r))$), with the value of $e_q^0$ half of that given in the expression. Again, $\rho_q^{j}$ is as given in Eq.~\ref{Eq:rhoqj}. 

\paragraph{Free energy} The free energy of the system in SQDFT is computed as
\begin{eqnarray}
\mathcal{F}(\bR) & = & h^3 \sum_{p=1}^{N_P} \sum_{q\in K_p} \bigg( \frac{2}{h^3} \sum_{j=0}^{n_{pl}} (\chi_q c_q^j + \zeta_q d_q^j) \rho_q^{j} +  \varepsilon_{xc}(\rho_{q}) \rho_{q} -   V_{xc}(\rho_p) \rho_q + \frac{1}{2}  (b_q - \rho_q) \phi_q  \nonumber \\
& &  + \frac{1}{2}  (\tilde{b}_q + b_q) V_{c,q}  - \frac{1}{2} \sum_{I \in D_{p}^b} \tilde{b}_{I,q} \tilde{V}_{I,q} + \frac{2 \sigma}{h^3} \sum_{j=0}^{n_{pl}} e_q^j \rho_q^{j} \bigg) \label{Eqn:FreeEnergy:FD} \,,
\end{eqnarray}
where the spatial integrals in Eq.~\ref{DM:groundstate} have been approximated using the trapezoidal rule in Eq.~\ref{Eqn:IntApprox} and $D_p^b$ is the set of all atoms (considering all atoms in $\R^3$) whose pseudocharges have overlap with the processor domain $\Omega_p$. We note that the exchange correlation energy $E_{xc}$ has been modeled using the Local Density Approximation (LDA) \cite{Kohn1965}, wherein $\epsilon_{xc}(\rho)$ is the sum of the exchange and correlation energy per particle of a uniform electron gas of density $\rho$. 

During the free energy calculation, the values of $c_q^j$ and $\rho^j_q$ determined as part of the electron density computation in the last SCF iteration are directly utilized. The Chebyshev coefficients $d_q^j$ and $e_q^j$ are determined using the discrete orthogonality of the Chebyshev polynomials \cite{gil2007numerical}. One \texttt{MPI\_Allreduce} command is utilized to simultaneously compute all the components of the energy. The computational cost per processor scales as $\mathcal{O}(n_{pl}N_d^p)$, which translates to an overall scaling of $\mathcal{O}(N)$ with respect to the number of atoms. 
%Even though the free energy needs to be computed only after determining the electronic ground state, we do so in every SCF iteration---a feature common to standard electronic structure codes---since the computational cost involved is negligible.  

%%%%%%%%%%%%%%%%%%%%%%%%%%%%%%%%%%%%%%%%%%%%%%%%%%%%%

\subsection{Atomic forces} \label{Subsec:Forces}
In order to update the positions of the atoms during the course of the QMD simulation, the Hellmann-Feynman forces on the nuclei (Eq.~\ref{Eqn:forces}) need to be computed.  
\paragraph{Local component}
The local component of the force has the following discrete form in parallel computations
\begin{equation}
\mathbf{f}_I^{l} = h^3 \sum_{p=1}^{N_P} \sum_{I' \in D_{p,I}^{b}} \sum_{q \in K_{p}} \bigg( \nabla_h b_{I'}\big|_q \phi_q + \frac{1}{2} \left(\tilde{b}_q+b_q\right) \nabla_h V_{c,I'}\big|_q + \frac{1}{2} \left(\nabla_h \tilde{b}_{I'}\big|_q + \nabla_h b_{I'}\big|_q \right) V_{c,q} \bigg) \label{Eqn:LF:Disc}
\end{equation}
where the integrals in Eq.~\ref{Eqn:forces} have been approximated using the trapezoidal rule (Eqn.~\ref{Eqn:IntApprox}) and $D_{p,I}^{b}$ is the set of the $I^{th}$ atom and its images whose pseudocharges have overlap with the processor domain $\Omega_p$. 

\paragraph{Nonlocal component}
The nonlocal component of the force in the SQ approach \cite{pratapa2016spectral} takes the following form in parallel computations
\begin{equation} \label{Eq:Discrete:NLforce}
\mathbf{f}_I^{nl} = -4 \sum_{p=1}^{N_P} \sum_{I' \in D_{p,I}^{c}} \sum_{q \in K_{p}} \bw_q^T  \mathbf{V}_{nl,q}^{I'} \nabla_{h,q} \left( \sum_{j=0}^{n_{pl}} c_q^j \mathbf{t}_q^{j} \right)
\end{equation}
where $D_{p,I}^{c}$ is the set of the $I^{th}$ atom and its images whose nonlocal projectors have overlap with the processor domain $\Omega_p$, $c_q^j$ is the coefficient of $T_j$ in the polynomial expansion of the Fermi-Dirac function (Eq.~\ref{Eq:cqj}); and $\mathbf{t}_q^{j}$ is determined using the recurrence relation:
\begin{eqnarray}
\mathbf{t}^{i+1}_q &=&  2\hat{\mathbf{H}}_q \mathbf{t}^{i}_q-\mathbf{t}^{i-1}_q \,, \quad i=1, 2, \ldots, n_{pl} \nonumber \\ 
\mathbf{t}_q^1 & = & \hat{\mathbf{H}}_q \mathbf{w}_q \,,\,\,  \mathbf{t}_q^0 = \mathbf{w}_q \,, \label{Eqn:Rec:FD1}
\end{eqnarray}
Note that unlike the iteration in Eq.~\ref{Eqn:Rec:FD} which proceeds up to $n_{pl}/2$, the above iteration proceeds up to $n_{pl}$ since some of the off-diagonal components of the density matrix are needed for the calculation of the nonlocal force in Eq.~\ref{Eq:Discrete:NLforce}, i.e., $\rho^j_q$ are not sufficient, rather $\mathbf{t}_q^j$ are required. 

\paragraph{Total atomic force}
The atomic force in SQDFT is then calculated as
\begin{eqnarray}
\mathbf{f}_I & = & h^3 \sum_{p=1}^{N_P} \sum_{I' \in D_{p,I}^{b}} \sum_{q \in K_{p}} \bigg( \nabla_h b_{I'}\big|_q \phi_q + \frac{1}{2} \left(\tilde{b}_q+b_q\right) \nabla_h V_{c,I'}\big|_q + \frac{1}{2} \left(\nabla_h \tilde{b}_{I'}\big|_q + \nabla_h b_{I'}\big|_q \right) V_{c,q} \bigg) \nonumber \\
& - & 4 \sum_{p=1}^{N_P} \sum_{I' \in D_{p,I}^{c}} \sum_{q \in K_{p}} \bw_q^T  \mathbf{V}_{nl,q}^{I'} \nabla_{h,q} \left( \sum_{j=0}^{n_{pl}} c_q^j \mathbf{t}_q^{j} \right) \,.
\end{eqnarray}

During the computation of the atomic forces, the values of $c_q^j$ that were determined as part of the electron density calculation during the last SCF iteration are directly utilized. As mentioned previously, the nodal gradient matrix  $\nabla_{h,q}$ is not generated/stored explicitly, rather its product with $\mathbf{t}_q^{j}$ is calculated in matrix-free fashion. The total atomic force on all atoms is simultaneously computed using a single \texttt{MPI\_Allreduce}. Since the storage per processor scales as $\mathcal{O}(3 N_s)$ and the computational effort per processor scales as $\mathcal{O}(n_{pl} N_s N_d^p) + \mathcal{O}(3 N_s)$\footnote{This corresponds to the  calculation of the nonlocal component of the force, which is the dominant cost in the atomic force calculation in SQDFT. Note that $\mathcal{O}(3 N_s)$ arises due to the three matrix-vector products arising in the multiplication with the finite-difference gradient.}, the overall  storage and computational cost scales as $\mathcal{O}(N)$ with respect to the number of atoms.
%%%%%%%%%%%%%%%%%%%%%%%%%%%%%%%%%%%%%%%%%%%%%%%%%%%%%%%%%%%%%%%%%%%%%%%%%%%%%%%%%%%%%%%%%%%%%%%%%%%%%%%%%%%%%%%%%%%%

\paragraph{Relation to classical Fermi Operator Expansion (FOE)}
The Clenshaw-Curtis SQ method bears some resemblance to the classical Fermi Operator Expansion (FOE) \cite{goedecker1995tight,goedecker1994efficient} in that both techniques use Chebyshev polynomials as the underlying basis for expanding the Fermi-Dirac function of a matrix. However, the key underlying difference is that the matrix in FOE corresponds to the Hamiltonian, whereas in SQ it corresponds to the nodal Hamiltonian. Therefore, truncation is automatically included within the SQ method and the key operation is reduced to local sparse matrix-vector products, as opposed to the global sparse matrix-matrix products in the FOE method. This 
% not only makes SQ much simpler to implement, but also 
makes SQ more efficient since (i) the Fermi level calculation does not require an outer loop, and (ii) inter-processor communication is needed just once per SCF iteration, unlike the FOE where it is required for every matrix-matrix multiplication. 
%In addition, it makes SQ more robust compared to FOE, since the effect of truncation on the Chebyshev expansion does not need to be accounted for in SQ. 
Finally, SQ also requires significantly less storage compared to FOE, making it especially well suited for modern high performance computing platforms.

%%%%%%%%%%%%%%%%%%%%%%%%%%%%%%%%%%%%%%%%%%%%%%%%%%%%%%%%%%%%%%%%%%%%%%%%%%%%%%%%%%%%%%%%%%%%%%%%%%%%%%%%%%%%%%%%%%%
\section{Results and discussion} \label{Sec:Results}
In this section, we demonstrate the accuracy, efficiency, and scaling of SQDFT in Kohn-Sham Density Functional Theory (DFT) calculations at high temperature. In all simulations, we employ a twelfth-order finite-difference discretization, norm-conserving Troullier-Martins pseudopotentials \cite{Troullier}, and the Local Density Approximation (LDA) \cite{Kohn1965} with the Perdew-Wang parametrization \cite{perdew1992accurate} of the correlation energy calculated by Ceperley-Alder \cite{Ceperley1980}. To ensure accuracy of the standard 3s3p pseudopotentials employed, temperatures were limited to $T \lesssim 80000$ K, where 2p states can be treated as fully occupied.\footnote{Calculations with deeper 2s2p3s3p pseudopotentials show 2p occupation of 5.9991 at 80000 K.
} 
To demonstrate applicability to metals and insulators alike, we consider two systems: (i) aluminum, a prototypical metal, and (ii) lithium hydride, a prototypical insulator. We compare the results obtained by SQDFT to benchmarks obtained by the finite-difference code SPARC \cite{ghosh2016sparc1,ghosh2016sparc2} and planewave code ABINIT \cite{ABINIT}, both of which solve the Kohn-Sham problem via diagonalization.

\subsection{Accuracy and convergence} \label{Subsec:AccuracyConvergence}
We first study the accuracy of SQDFT, i.e., we verify the convergence of computed energies and atomic forces with respect to key SQ parameters (i.e., quadrature order $n_{pl}$ and truncation radius $R_\text{cut}$) as well as spatial discretization (i.e., mesh-size $h$). As representative systems, we consider a $32$-atom cell of aluminum at the equilibrium lattice constant of $7.78$ Bohr and a $64$-atom cell of lithium hydride at the equilibrium lattice constant of $7.37$ Bohr, with all atoms randomly displaced by up to $15 \%$ of the equilibrium interatomic distance. We select the electronic temperature/smearing to be  $\sigma=4$ eV.

\begin{figure}[h]
\centering
\subfloat[Convergence with respect to $n_{pl}$]{\label{Conv_npl}\includegraphics[width=0.45\textwidth]{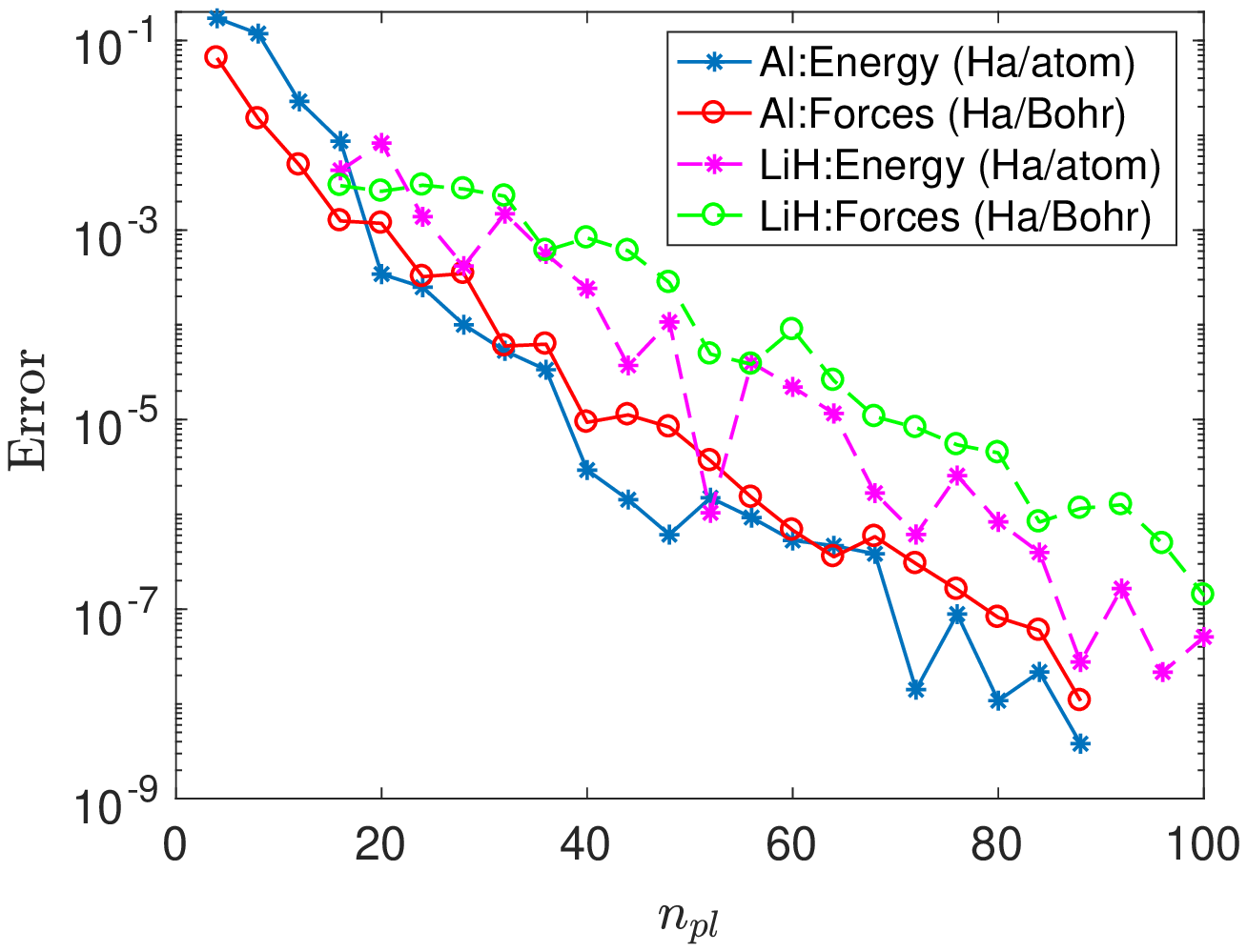} } 
\qquad
\subfloat[Convergence with respect to $R_\text{cut}$]{\label{Conv_rcut}\includegraphics[width=0.45\textwidth]{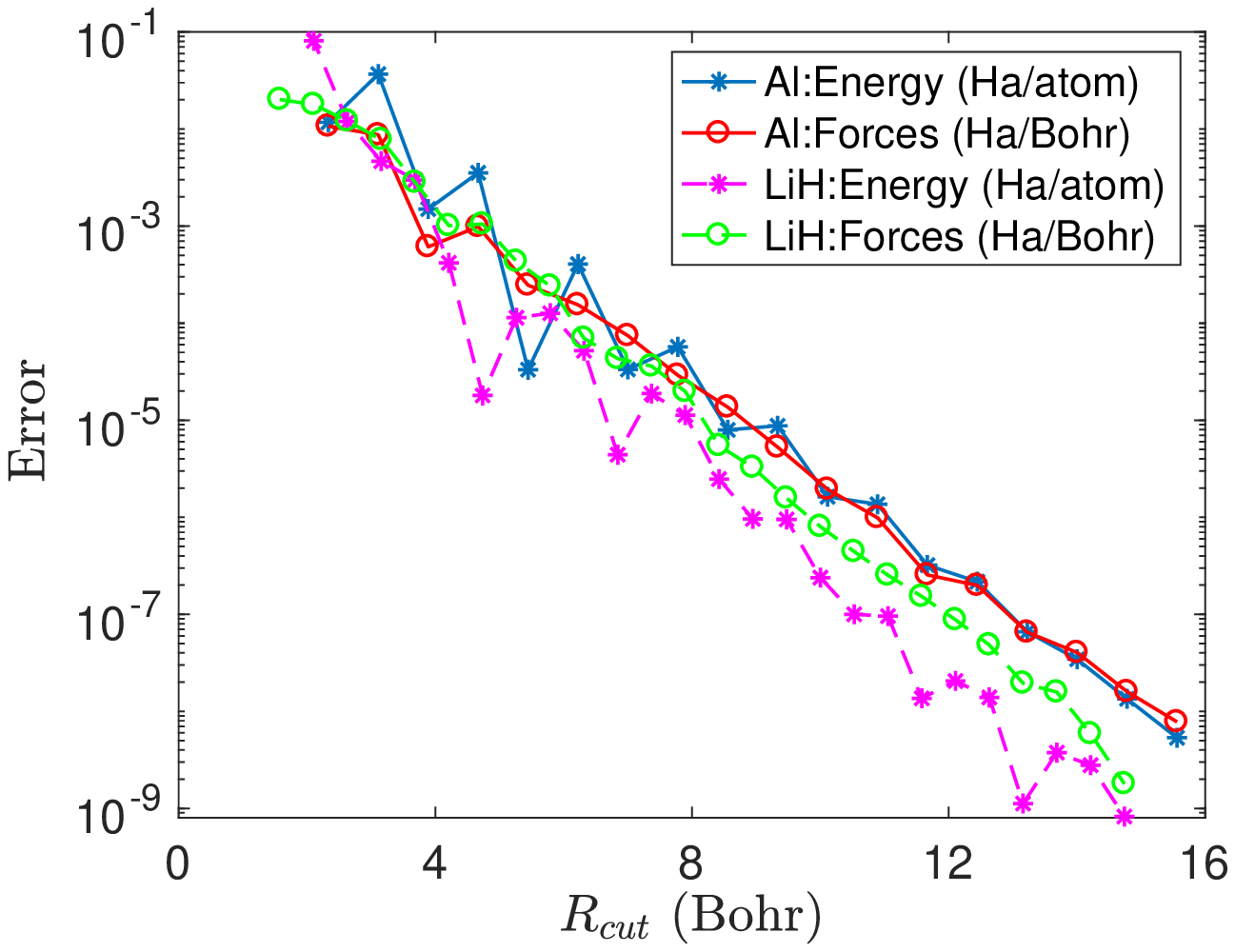} } 
\caption{Convergence of energy and forces in SQDFT with respect to quadrature order $n_{pl}$ and truncation radius $R_\text{cut}$ for aluminum and lithium hydride systems. The error in energy is the magnitude of the difference and the error in forces is the maximum difference in any component, with diagonalization result obtained by finite-difference code SPARC at the same mesh-size as reference.}
\label{Fig:Conv_npl_rcut}
\end{figure}

First, we verify the convergence of SQDFT energies and forces with respect to $n_{pl}$ and $R_\text{cut}$ in Fig.~\ref{Fig:Conv_npl_rcut}, with the reference diagonalization answers obtained by SPARC at the same mesh-size and a $4 \times 4 \times 4$ Monkhorst-Pack grid for Brillouin zone integration.\footnote{Unlike standard codes, SQDFT can obtain the infinite crystal result without recourse to Brillouin zone integration.} We choose mesh-sizes of $h=0.7780$ and $h=0.5264$ Bohr for the aluminum and lithium hydride systems, respectively. It is clear that SQDFT obtains exponential convergence in the energy and atomic forces with respect to both parameters, in agreement with previous studies \cite{suryanarayana2013spectral,pratapa2016spectral,suryanarayana2017nearsightedness}. In particular, $\{ n_{pl},R_\text{cut} \} \sim \{28,6\}$ and $\{ n_{pl},R_\text{cut} \} \sim \{40,6\}$ are sufficient to obtain chemical accuracy in both the energy and forces for the aluminum and lithium hydride systems, respectively. Importantly, these values further reduce as the smearing/temperature is increased \cite{pratapa2016spectral}, which makes SQDFT particularly attractive for high-temperature simulations. Note that neither the energy nor the atomic forces are variational with respect to $n_{pl}$ and $R_\text{cut}$, hence the non-monotonic convergence in Fig.~\ref{Fig:Conv_npl_rcut}. 

\begin{figure}[h]
\centering
\subfloat[Aluminum]{\label{mesh_Al}\includegraphics[width=0.45\textwidth]{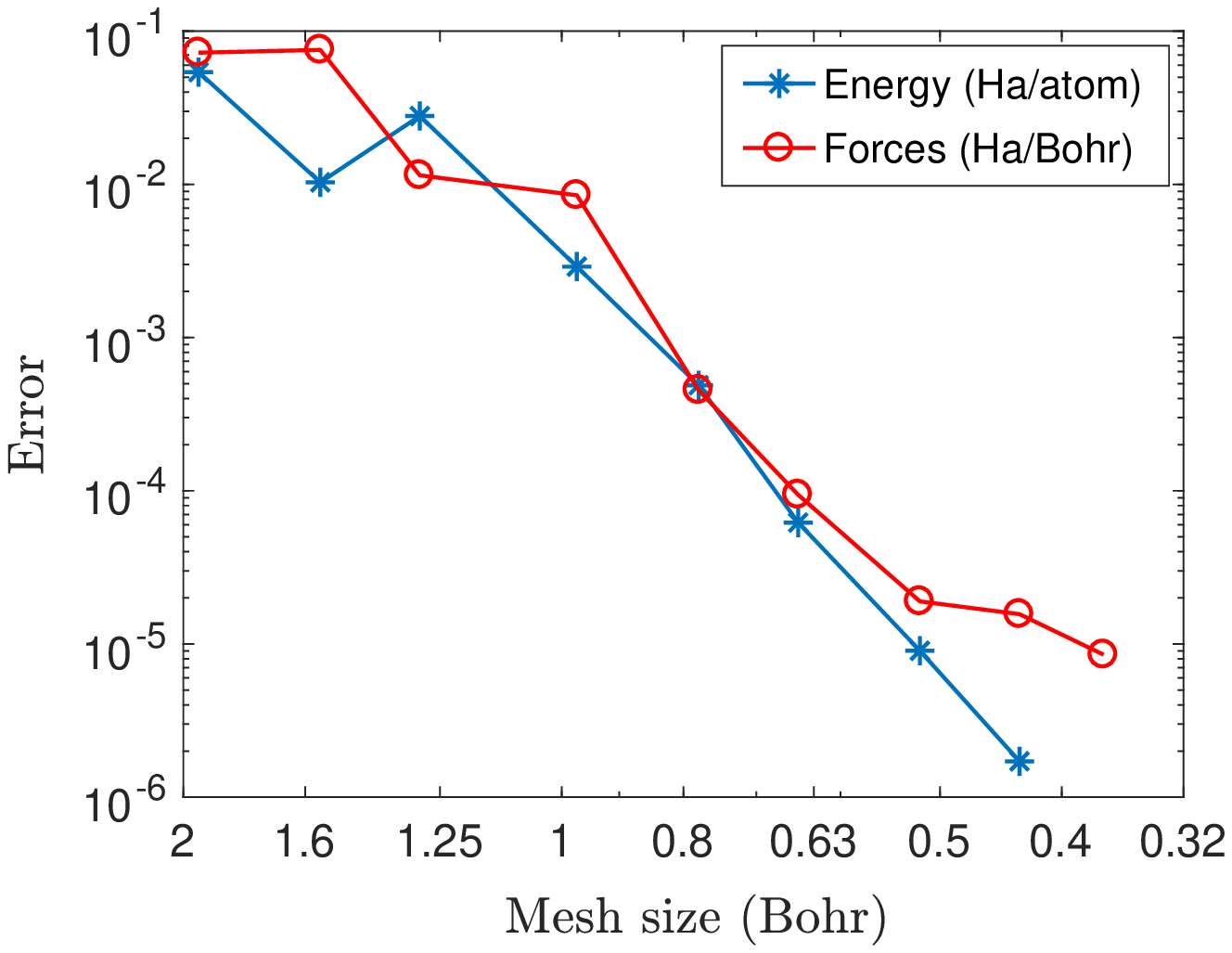} } 
\qquad
\subfloat[Lithium hydride]{\label{mesh_LiH}\includegraphics[width=0.45\textwidth]{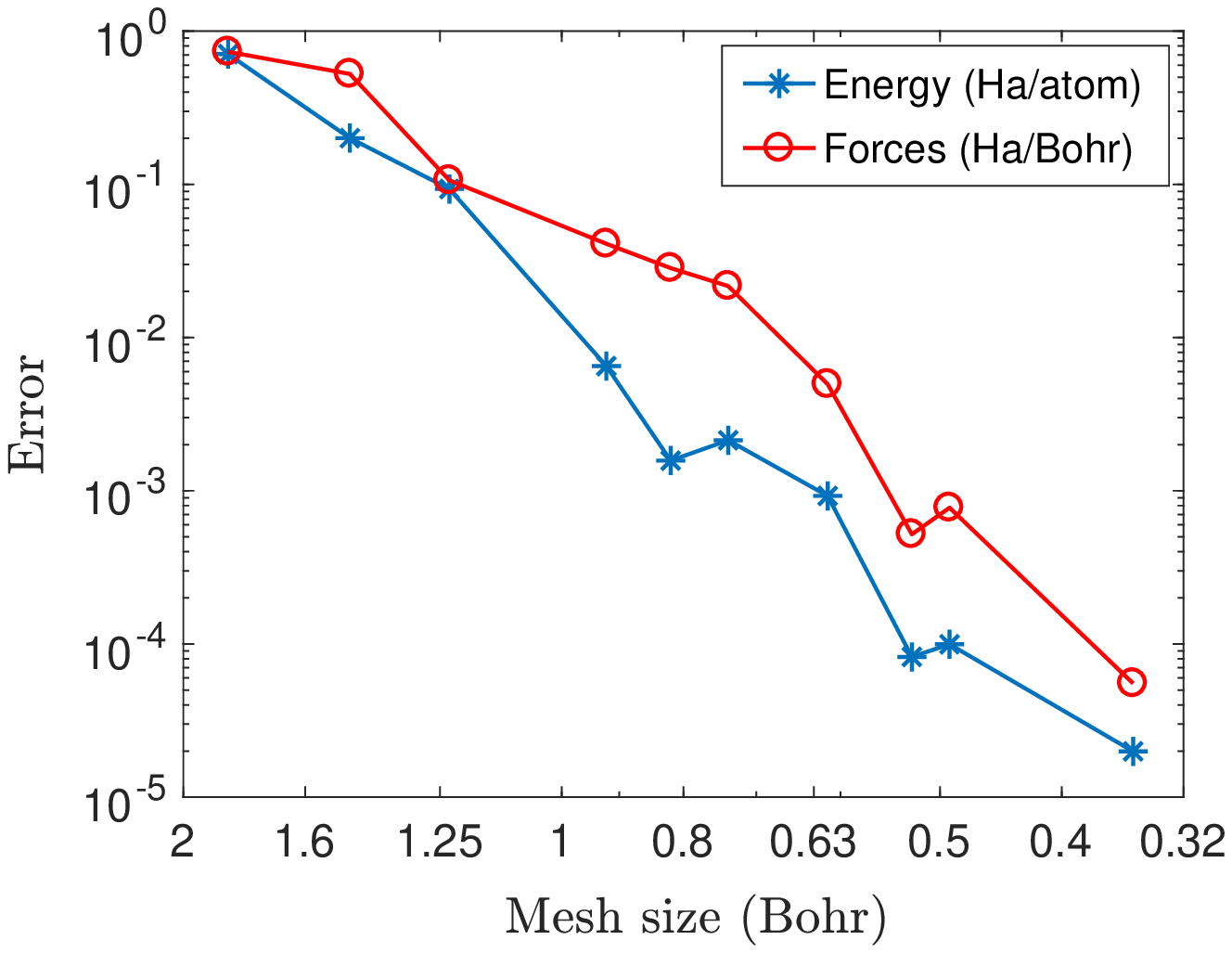} } 
\caption{Convergence of energy and forces in SQDFT with respect to mesh-size $h$ for aluminum and lithium hydride systems. The error in energy is the magnitude of the difference and error in forces is the maximum difference in any component, with results obtained by planewave code ABINIT as reference.}
\label{Fig:mesh_convergence}
\end{figure}

Next, we verify the convergence of SQDFT energies and forces with mesh-size $h$ to those computed by the established  planewave code ABINIT. To do so, we utilize $n_{pl}=160$ and $R_\text{cut} = 10$ Bohr in SQDFT, which are sufficient to put the associated errors well below the mesh errors of interest (see Fig.~\ref{Fig:Conv_npl_rcut}). In ABINIT, we employ a planewave cutoff of $50$ Ha and a $4 \times 4 \times 4$ Monkhorst-Pack grid for Brillouin zone integration, which results in energy and forces that are converged to within $10^{-6}$ Ha/atom and $10^{-6}$ Ha/Bohr, respectively. As shown in Fig.~\ref{Fig:mesh_convergence}, both the energy and atomic forces in SQDFT converge rapidly and systematically, with chemical accuracy readily obtained. Notably, we see that energies and forces converge at comparable rates, without the need for additional measures such as double-grid \cite{OnoHir99} or high-order integration \cite{BobSchChe15}. Therefore, accurate forces are easily obtained, as needed for structural relaxations and molecular dynamics simulations.

%%%%%%%%%%%%%%%%%%%%%%%%%%%%%%%%%%%%%%%%%%%%%
\subsection{Scaling and performance} \label{Subsec:ParallelScaling}
We now study the scaling and performance of SQDFT on large-scale parallel computers with up to tens of thousands of processors. Specifically, we investigate the strong and weak scaling of SQDFT for aluminum and lithium hydride systems on the \texttt{Sierra} and \texttt{Quartz} supercomputers at the  Lawrence Livermore National Laboratory (LLNL) \cite{LLNLwebMachines}. In all calculations, we employ a smearing of $\sigma=4$ eV and utilize (i) $h=0.7780$ Bohr and $\{ n_{pl},R_\text{cut} \} = \{28,6.224 \text{ Bohr}\}$ for the aluminum systems, and (ii) $h=0.5264$ Bohr and $\{ n_{pl},R_\text{cut} \} = \{40,6.387 \text{ Bohr}\}$ for the lithium hydride systems. These parameters are sufficient to obtain chemical accuracy of $0.001$ Ha/atom and $0.001$ Ha/Bohr in the energy and atomic forces, respectively, as demonstrated in the previous section.  

\begin{figure}[h]
\centering
\subfloat[Strong scaling]{\label{strong_scale}\includegraphics[width=0.45\textwidth]{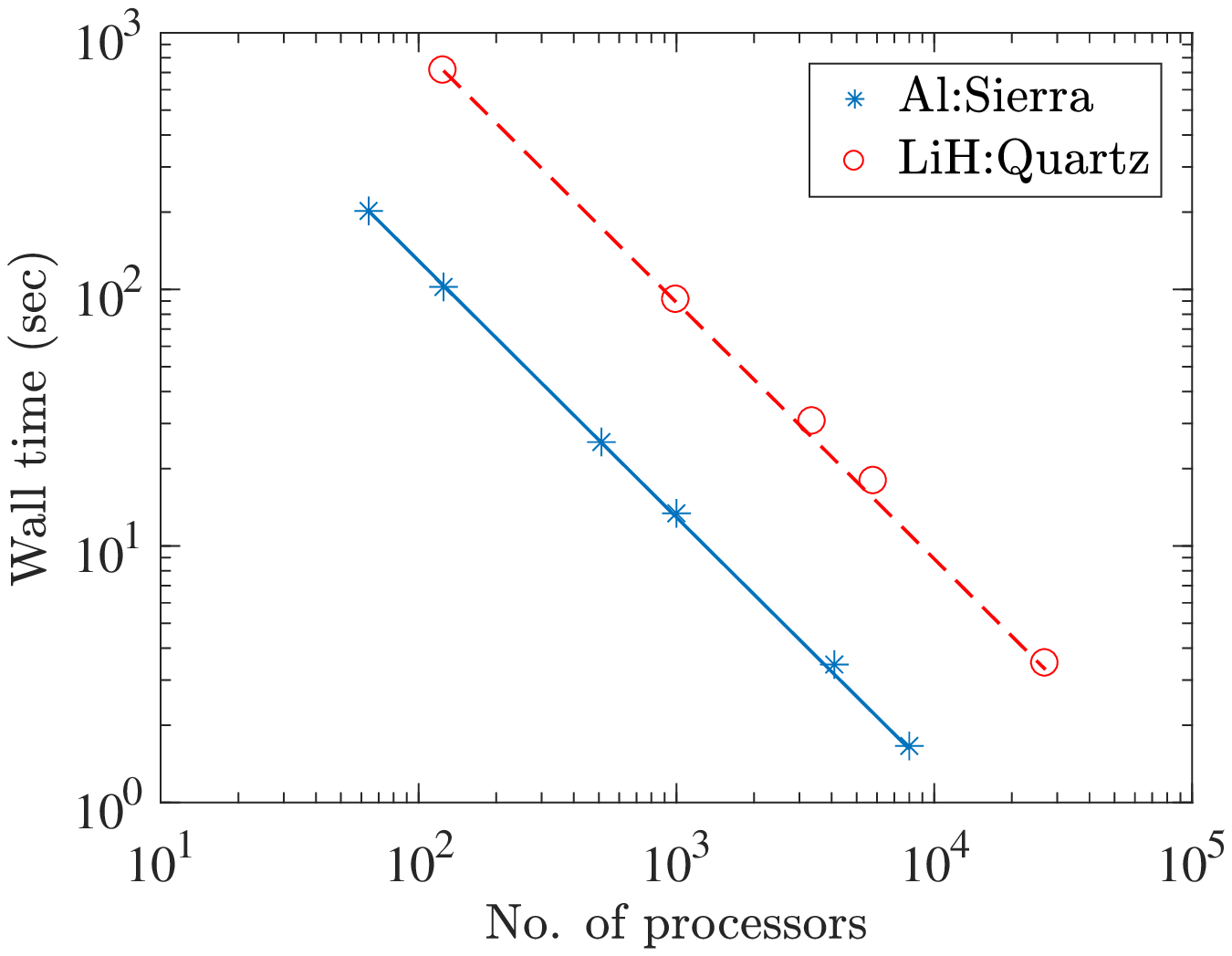} } 
\qquad
\subfloat[Weak scaling]{\label{weak_scale}\includegraphics[width=0.45\textwidth]{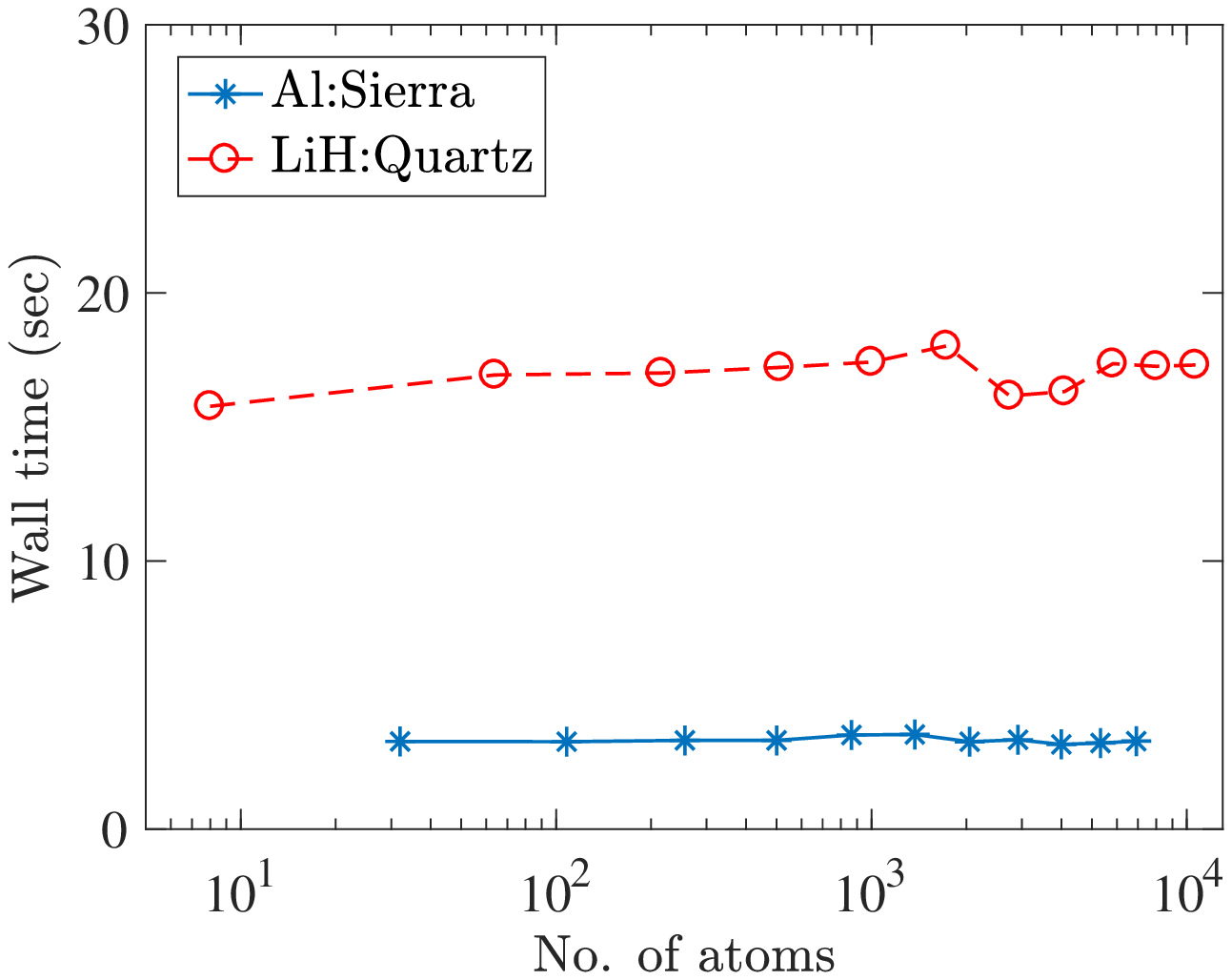} } 
\caption{Parallel scaling of a single SCF iteration in SQDFT on \texttt{Sierra} and \texttt{Quartz} supercomputers at the  Lawrence Livermore National Laboratory (LLNL) \cite{LLNLwebMachines}. In the strong scaling plot, the straight lines represent ideal scaling. All parameters have been chosen so as to achieve chemical accuracy of $0.001$ Ha/atom and $0.001$ Ha/Bohr in energy and atomic forces, respectively.}
\label{Fig:Parallel_scaling}
\end{figure}

First, we perform a strong scaling study for a $2048$-atom aluminum system on \texttt{Sierra} and a $1728$-atom lithium hydride system on \texttt{Quartz}, with atoms randomly displaced in both systems. For aluminum, the number of processors on \texttt{Sierra} is varied from $64$ to $8000$. For lithium hydride, the number of processors on \texttt{Quartz} is varied from $125$ to $27000$. The wall times per SCF iteration so obtained are presented in Fig.~\ref{strong_scale}. Relative to the smallest number of processors, on the largest number of processors SQDFT achieves $97 \%$ parallel efficiency for aluminum on \texttt{Sierra} and $95 \%$ for lithium hydride on \texttt{Quartz}. It is clear that SQDFT demonstrates excellent strong scaling. Notably, the wall time per SCF iteration for systems containing $\sim 2000$ atoms can be reduced to less than $5$ seconds.

Next, we perform a weak scaling study for aluminum and lithium hydride. For aluminum on \texttt{Sierra}, we increase the system size from $32$ to $6912$ atoms, while increasing the number of processors from $64$ to $13824$, maintaining two processors per atom for all systems. For lithium hydride on \texttt{Quartz}, we increase the system size from $8$ to $10648$ atoms, while increasing the number of processors from $27$ to $35937$, maintaining $\sim$ 3.4 processors per atom for all systems. The systems are generated by replicating $4$-atom and $8$-atom unit cells of aluminum and lithium hydride, respectively, with one atom in each unit cell randomly perturbed. We present the results so obtained in Fig.~\ref{weak_scale}. We find the scaling with system size for aluminum on \texttt{Sierra} to be $\mathcal{O}(N^{1.00})$ and the scaling for lithium hydride on \texttt{Quartz} to be $\mathcal{O}(N^{1.01})$.
%\footnote{Since all the systems are obtained by periodically replicating a single unit cell, the number of SCF iterations remains constant with system size. Indeed one of the challenges in attaining $\mathcal{O}(N)$ scaling for metallic systems is rendering the number of SCF iterations to be independent of system size. This is significantly mitigated in high-temperature calculations, wherein the large value of smearing helps in accelerating SCF convergence \cite{lin2013elliptic}.} 
It is clear that SQDFT demonstrates excellent weak scaling, with near perfect $\mathcal{O}(N)$ scaling with respect to system size in practical calculations. 

Overall, the excellent strong and weak scaling of SQDFT up to tens of thousands of processors makes it possible to perform high-temperature Kohn-Sham molecular dynamics simulations at large length and time scales, as we show below.   

%%%%%%%%%%%%%%%%%%%%%%%%%%%%%%%%%%%%%%%%%%%%%
\subsection{High-temperature quantum molecular dynamics} \label{Subsec:QMD}
We now consider the accuracy and efficiency of SQDFT in high-temperature Born-Oppenheimer quantum molecular dynamics (QMD) simulations. Such simulations are a cornerstone of modern warm dense matter theory \cite{gradesred2014}, providing equation-of-state and shock-compression predictions of unprecedented accuracy, up to temperatures of $\sim 100$ eV and pressures of 100s of Mbar; see, e.g.,  \cite{humilkre2016,zhadrimil2016,zhadrimil2017b,shazurham2017,drisoumil2017,zhadrimil2017}. At temperatures above $\sim 100$ eV, conventional Kohn-Sham methods become prohibitively expensive, and so alternative methods such as OFMD \cite{lamclegil2006} and, more recently, path integral Monte Carlo (PIMC) \cite{zhadrimil2017} have been employed to reach temperatures of 1000s of eV and higher. With sufficiently deep potentials, however, the new SQ methodology makes possible Kohn-Sham MD at temperatures of 1000s of eV as well.

%\footnote{SQDFT is able to perform both NVE and NVT QMD simulations. Though we focus on NVE calculations in this work, we have verified the ability of SQDFT to conserve both the total energy (system plus thermostat) and temperature in NVT QMD simulations.}. 
As a representative example, we choose an $864$-atom aluminum system and perform a $0.15$ ps NVE QMD simulation with time step of $0.1$ fs.\footnote{Due to the high velocities of the ions in high-temperature simulations, the time step has been chosen significantly smaller than in ambient calculations \cite{zhang2016extended}.} We use a mesh-size $h=0.7780$ Bohr, quadrature order $n_{pl}=28$, truncation radius $R_\text{cut}=6.224$ Bohr, initial ionic temperature $T = 116045$ K, initial atomic positions close to perfect FCC crystal with one atom in each 4-atom unit cell randomly displaced by the same amount, and initial velocities randomly assigned based on the Maxwell-Boltzmann distribution. We integrate the equations of motion using the Leapfrog method \cite{rapaport2004art}. The values of $n_{pl}$ and $R_\text{cut}$ have been chosen so as to put the associated errors close to an order of magnitude lower than the discretization error, which is $\sim 0.001$ Ha/atom and $\sim 0.001$ Ha/Bohr in the energy and forces, respectively. At each MD step, we set the electronic temperature equal to the ionic temperature, e.g, $\sigma=10$ eV at the start of the simulation. 

\begin{figure}[h]
\centering
\subfloat[Total energy]{\label{MD:TE}\includegraphics[width=0.48\textwidth]{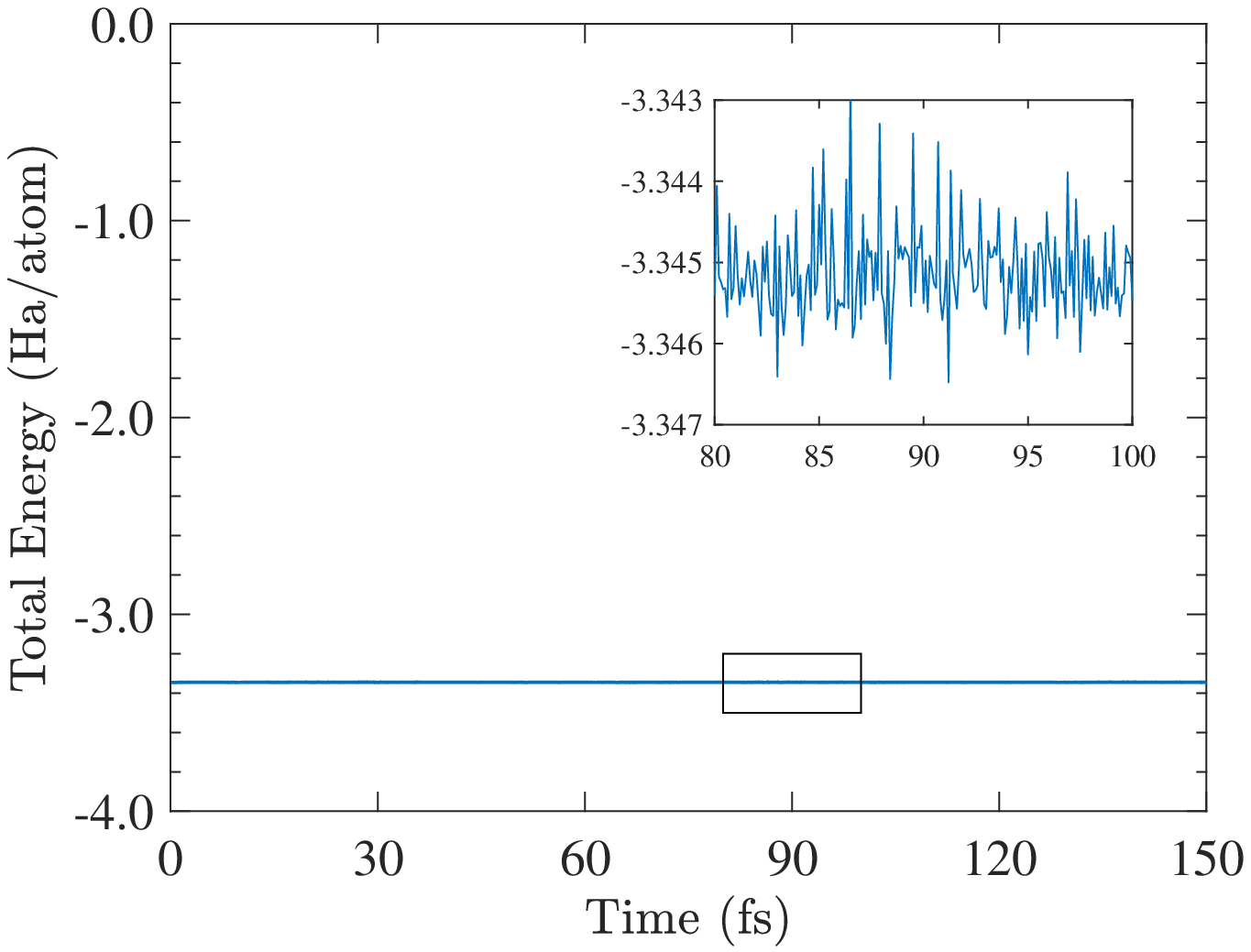} } 
\qquad
\subfloat[Temperature]{\label{MD:Temp}\includegraphics[width=0.45\textwidth]{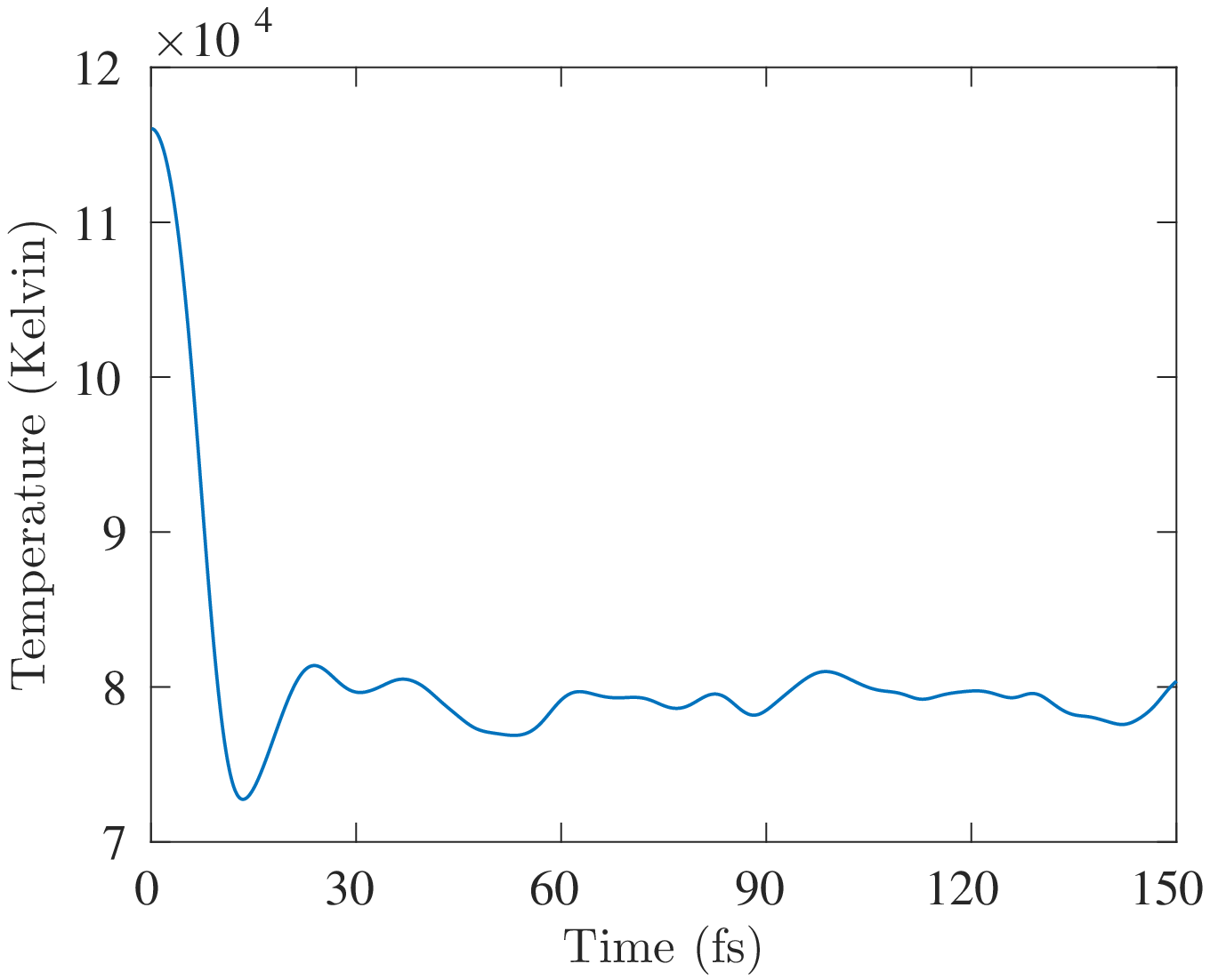} } 
\caption{Variation of total energy and temperature during $864$-atom aluminum NVE QMD simulation.}
\label{Fig:MD_NVE}
\end{figure}

\begin{figure}[h]
\centering
\includegraphics[width=0.43\textwidth]{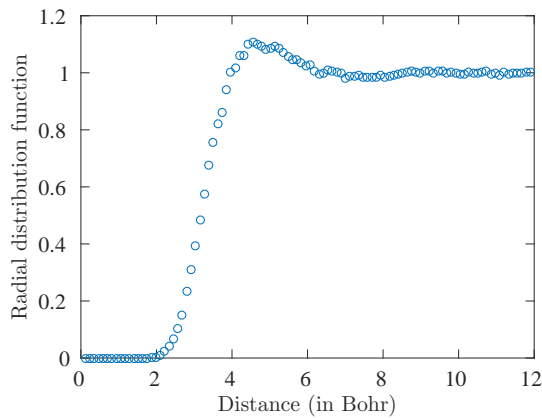} 
\caption{Radial distribution function for $864$-atom aluminum NVE QMD simulation.}
\label{Fig:MD_RDF}
\end{figure}

We perform the simulation on $3375$ processors on \texttt{Quartz} to obtain a wall clock time of $\sim 30$ seconds per QMD step\footnote{Since the computational cost of SQDFT reduces with temperature, the wall time will further reduce as the temperature is increased.}. In Fig. \ref{Fig:MD_NVE}, we plot the variation of  the total energy and temperature of the system over the course of the simulation. We observe that the temperature settles after $\sim$ $30$ fs, subsequent to which the mean and standard deviation of the total energy are $-3.3451$ and $4.7 \times 10^{-4}$ Ha/atom, respectively. In addition, the drift in total energy as obtained from a linear fit is $\sim$ $2.7 \times 10^{-4}$ Ha/atom-ps. 
%\footnote{The drift in total energy further reduces as the spatial discretization is refined.}  
%The small standard deviation in the total energy and the lack of any significant drift verifies that there are no systematic errors in SQDFT, which further confirms the accuracy of the energy and atomic forces in SQDFT. 
SQDFT thus shows excellent energy conservation, consistent with the accurate atomic forces obtained. 
We also plot the calculated radial distribution function 
%---factor that signifies change in atomic density relative to the average atomic density--- 
in Fig.~\ref{Fig:MD_RDF}, which is in agreement with previous studies \cite{white2013orbital}. 
%Overall, these results demonstrate the accuracy and efficiency of performing high-temperature QMD simulations using SQDFT. 

%%%%%%%%%%%%%%%%%%%%%%%%%%%%%%%%%%%%%%%%%%%%%%%%%%%%%%%%%%%%%%%%%%%%%%%%%%%%%%%%%%%%%%%%%%%%%%%%%%%%%%%%%%%%%%%%%%%%%
\section{Concluding remarks} \label{Sec:Conclusions}
%We have presented SQDFT: an $\mathcal{O}(N)$ framework for performing large-scale Kohn-Sham Density Functional Theory (DFT) calculations at high temperature. Specifically, we have developed a parallel finite-difference implementation of the infinite-cell Clenshaw-Curtis Spectral Quadrature (SQ) method, in which results for the infinite crystal are obtained by expressing quantities of interest as bilinear forms or sums of bilinear forms, that are then approximated by spatially localized Clenshaw-Curtis quadrature rules. We have verified the accuracy of SQDFT by checking convergence of energy and atomic forces with respect to the SQ parameters to results obtained by diagonalization; and their convergence with spatial discretization to well-established planewave results. We have shown that SQDFT achieves excellent strong and weak scaling on large-scale parallel computer systems consisting of tens of thousands of processors, having near perfect $\mathcal{O}(N)$ scaling with system size and achieving wall times as low as a few seconds per SCF iteration. We have also demonstrated that SQDFT is able to conserve the total energy in NVE quantum molecular dynamics (QMD) simulations. 

We presented SQDFT: a large-scale parallel implementation of the Spectral Quadrature (SQ) method for $\mathcal{O}(N)$ Kohn-Sham density functional theory calculations at high temperature.
Specifically, we developed an efficient and scalable finite-difference implementation of the infinite-cell Clenshaw-Curtis SQ approach, in which results for the infinite crystal are obtained by expressing quantities of interest as bilinear forms or sums of bilinear forms, that are then approximated by spatially localized Clenshaw-Curtis quadrature rules. We demonstrated the accuracy of SQDFT by showing systematic convergence of energies and atomic forces with respect to quadrature order and truncation radius to reference diagonalization results, and convergence with mesh spacing to established planewave results, for both metallic and insulating systems. In all cases, chemical accuracy was readily obtained. We demonstrated excellent strong and weak parallel scaling on computer systems consisting of tens of thousands of processors, with near perfect $\mathcal{O}(N)$ scaling with system size, and wall clock times as low as a few seconds per SCF iteration for insulating and metallic systems of $\sim$ 2000 atoms. Finally, we verified the accuracy and efficiency of SQDFT in large-scale quantum molecular dynamics (QMD) simulations at high temperature, demonstrating excellent energy conservation and QMD step times of $\sim$ 30 seconds for an 864-atom aluminum system at $\sim 80000$ K.

In the present work, we have focused on high-temperature Kohn-Sham DFT calculations. 
%However, the SQ method does not make any assumption  about the value of the temperature and is in fact applicable to any finite temperature. 
However, the SQ method is applicable at lower temperatures as well, with larger prefactor, as we show in Appendix \ref{App:LowT}. A possible approach to reduce this prefactor is to generate a localized orthonormal reduced basis (e.g., \cite{lin2012adaptive,zhang2017adaptive}), subsequent to which the SQ method is applied to the finite-difference Hamiltonian projected into this basis. This is indeed a promising path to $\mathcal{O}(N)$ DFT calculations of metals and insulators at ambient conditions which the authors are pursuing presently.

%%%%%%%%%%%%%%%%%%%%%%%%%%%%%%%%%%%%%%%%%%%%%%%%%%%%%%%%%%%%%%%%%%%%%%%%%%%%%%%%%%%%%%%%%%%%%%%%%%%%%%%%%%%%%%%%%%%%%%
%%%%%%%%%%%%%%%%%%%%%%%%%%%%%%%%%%%%%%%%%%%%%%%%%%%%%%%%%%%%%%%%%%%%%%%%%%%%%%%%%%%%%%%%%%%%%%%%%%%%%%%%%%%%%%%%%%%%%%

\section*{Acknowledgements}
This work was supported in part by the National Science Foundation (Grant number 1333500), and performed in part under the auspices of the U.S. Department of Energy by Lawrence Livermore National Laboratory under Contract DE-AC52-07NA27344. Early support from the Exascale Co-design Center for Materials in Extreme Environments supported by Office of Science Advanced Scientific Computing Research Program, and subsequent support from the Laboratory Directed Research and Development program at the Lawrence Livermore National Laboratory is gratefully acknowledged.
%%%%%%%%%%%%%%%%%%%%%%%%%%%%%%%%%%%%%%%%%%%%%%%%%%%%%%%%%%%%%%%%%%%%%%%%%%%%%%%%%%%%%%%%%%%%%%%%%%%%%%%%%%%%%%%%%%%%%%
%%%%%%%%%%%%%%%%%%%%%%%%%%%%%%%%%%%%%%%%%%%%%%%%%%%%%%%%%%%%%%%%%%%%%%%%%%%%%%%%%%%%%%%%%%%%%%%%%%%%%%%%%%%%%%%%%%%%%%

\appendix

\vspace{10mm}

{\LARGE \bf Appendix}

%%%%%%%%%%%%%%%%%%%%%%%%%%%%%%%%%%%%%%%%%%%%%%%%%%%%%%%%%%%%%%%%%%%%%%%%%%%%%%%%%%%%%%%%%%%%%%%%%%%%%%%%%%%%%%%%%%

\section{Ambient temperature Kohn-Sham calculations} \label{App:LowT}
\begin{figure}[h]
\centering
\includegraphics[width=0.49\textwidth]{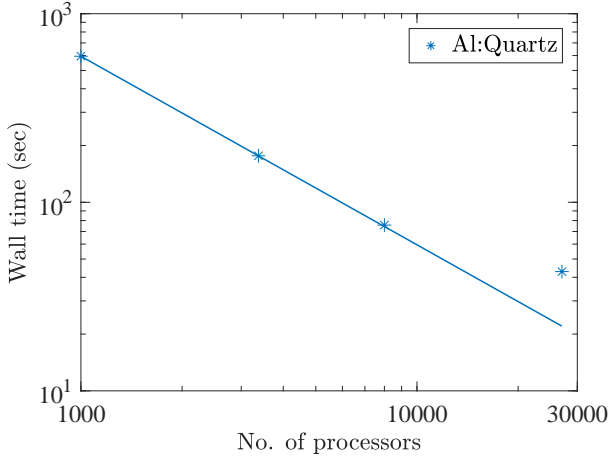}  
\caption{Strong scaling of single SCF iteration in SQDFT for $864$-atom aluminum system at $\sigma = 0.27$ eV on \texttt{Quartz} supercomputer at the  Lawrence Livermore National Laboratory (LLNL) \cite{LLNLwebMachines}. The straight line represents ideal scaling. All parameters have been chosen so as to achieve chemical accuracy of $0.001$ Ha/atom and $0.001$ Ha/Bohr in energy and atomic forces, respectively.}
\label{Fig:Parallel_scaling_lowT}
\end{figure}

Though the focus of the present work has been Kohn-Sham calculations at high temperature, SQDFT can also be utilized at ambient temperature, albeit with a larger prefactor. This increase in prefactor may, however, be mitigated by the excellent parallel scaling of SQDFT on large-scale parallel computers. 
In order to demonstrate this, we consider an $864$-atom randomly perturbed aluminum system with smearing $\sigma = 0.27$ eV, as typical in calculations of metallic systems at ambient conditions in order to facilitate self-consistent convergence \cite{VASP,ABINIT}. We utilize $h=0.7780$ Bohr and $\{ n_{pl},R_\text{cut} \} = \{320,18.672 \text{ Bohr} \}$, which are sufficient to obtain chemical accuracy of $0.001$ Ha/atom and $0.001$ Ha/Bohr in the energy and atomic forces, respectively. We perform the calculations on \texttt{Quartz}, where the number of processors is varied from $1000$ to $27000$, the results of which are presented in Fig.~\ref{Fig:Parallel_scaling_lowT}. Relative to $1000$ processors, the efficiency of SQDFT on $8000$ processors is larger than $98 \%$, but on $27000$ processors, the efficiency drops to $51 \%$. 
%The reduced efficiency on $27000$ processors can be attributed to the significant communication required for the nodal Hamiltonians relative to the computational work per processor. 
The reduced efficiency at this temperature at the largest processor counts arises due to the increased communications required for the larger nodal Hamiltonians ($R_\text{cut} = 18.672 \text{ Bohr}$) relative to the computational work per processor. 
However, SQDFT is still able to achieve wall times of less than a minute per SCF iteration, which demonstrates its ability to perform large-scale Kohn-Sham quantum molecular dynamics simulations (QMD) even at ambient temperature, given sufficient number of processors.

% \clearpage
\bibliographystyle{ReferenceStyle}
%\bibliography{Manuscript}

\end{document}